**Quantum Reciprocity: A Structured-Bath Hamiltonian for Coherent Amplification**

Ridwan Sakidja
Department of Physics, Astronomy & Materials Science, Missouri State University, Springfield, MO 65897, USA

**Abstract**

Macroscopic quantum amplifiers maintain coherence even while strongly coupled to their surroundings, demonstrating that coherence can be preserved through architecture rather than isolation. Here we derive a finite structured-bath Hamiltonian in which dissipation and feedback originate from the same microscopic couplings. The resulting self-energy $\Sigma(\omega)$ exhibits coupled real and imaginary parts whose evolution reproduces the breathing dynamics observed in Josephson quantum amplifiers. This establishes quantum reciprocity: macroscopic coherence lives not in isolation, but in structured connection. We numerically validate this principle by engineering a six-qubit structured bath to demonstrate controllable transitions from dissipation to amplification. This architectural core serves as the foundation for a proposed multi-scale workflow to transform quantum noise into a design resource, preserving coherence not through isolation but through architectural reciprocity.

**I. Introduction**

The 2025 Nobel Prize in Physics awarded to John Clarke, Michel Devoret, and John Martinis recognized the realization of Josephson-based parametric amplifiers capable of quantum-limited gain while preserving phase coherence[1], [2]. This achievement resolved the paradox of measurement without collapse, proving that a quantum state can be revealed without being destroyed[3]. It also marked a turning point in macroscopic quantum engineering, confirming that coherence can be sustained not by isolation, but through architectural control. Recent experiments at Chalmers University reinforce this logic [4], [5]. By synchronizing amplifier activation with qubit readout, they demonstrated that coherence and gain can be preserved while reducing back-action and thermal load by over ninety percent, all without compromising quantum-limited performance. These developments reflect principles formalized in this framework, where coherence, feedback, and memory emerge from structured coupling rather than fitted reconstruction.

At the heart of this discovery lies a deeper principle: macroscopic quantum coherence is sustained not by isolation but by a carefully engineered exchange between system and environment. The amplifier shapes its surroundings into a reciprocal partner that allows coherence to circulate rather than dissipate. This reframes the conventional division between system and bath. If the environment is architected rather than assumed infinite, it becomes an active participant in coherence.

This raises a central question: How can the flow of coherent information be managed through the environment, and how can this reciprocity be sustained and designed? Recent experiments confirm that coherence can be preserved through structured coupling rather than isolation, validating this architectural logic. In systems with strong internal correlation, especially those involving structured bath interfaces, a quantized treatment is necessary. Classical approximations met with a significant challenge nit because they are numerically weak but because they are conceptually incoherent in representing shared quantum agency. When dissipation and feedback arise from the same microscopic couplings, coherence becomes a structural feature, not a statistical anomaly. This mandates a first-principles framework that preserves the physical origin of both gain and loss.

**Different representations of open quantum systems**
Modern treatments of non-Markovian dynamics, including Hierarchical Equations of Motion (HEOM) [6], [7] and tensor-network methods (TEMPO) [8], are powerful and well-validated. They reconstruct the bath influence by approximating the correlation function as a sum of effective decay channels or virtual sites. This reconstructive approach is computationally efficient and numerically exact within its scope. However, it operates at the level of the spectral density $J(\omega)$: given an empirical or calculated spectrum, these methods infer the underlying dynamics.

The Energy Participation Ratio (EPR) framework takes a complementary geometric approach. EPR solves the static electromagnetic problem, i.e. how energy is distributed among circuit elements for each classical mode. It provides the spatial and energetic foundation of the device but, by construction, captures only the real (Hermitian) part of the self-energy; the temporal evolution and dissipative/reactive feedback are absent.

The structured-bath framework introduced here represents a distinct form of modeling. It derives the complex self-energy directly from a finite Hamiltonian with explicitly defined, physically meaningful couplings. Instead of inferring environmental effects from spectral densities or participation ratios, the approach begins with the architecture itself and predicts how coherence and dissipation emerge from its structure.

Within this framework, EPR provides the geometric foundation by identifying which elements participate in which modes and to what extent. HEOM and TEMPO describe how a given spectral density governs the dynamical exchange between system and bath. The structured-bath framework complements these views by taking the inverse route: given a desired dynamical behavior, it prescribes the coupling configuration that realizes it.

These perspectives are naturally connected. EPR participation maps can inform the spatial layout of a structured bath; the resulting predictions can be examined with HEOM or TEMPO to confirm temporal behavior; and such dynamical checks can validate that the engineered near field produces the intended self-energy response.

**The present approach**
Following our recent work[9], [10], where we introduced the concept of a structured, quantized environment as a finite intermediary between system and reservoir, we now extend this principle toward active control. In our previous study, we established that a structured bath could act as an effective interface between a quantum system and the outer continuum, serving as a quantized buffer that mediates coherence flow and information exchange. We further demonstrated that the parameters of this structured bath can be reverse-engineered solely from the dynamics of the central qubit, revealing the hidden topology and coupling strengths of the near field.

Another finding from that work was that memory can persist even under weak coupling, provided that thermal dissipation is sufficiently suppressed. This leads to the formation of coherent memory traps within the layered bath, where correlations are retained and exchanged in a controllable manner. In the present work, we build directly upon these insights by treating the structured bath not only as a passive memory element but as an active quantum amplifier. By engineering the topology and coupling parameters of the finite bath network, we show that the same quantized architecture can be tuned to exhibit gain, delay, or amplification, thereby functioning as a dynamic mediator of energy and information flow. The model is exactly solvable and incorporates a local coupling to an external thermal reservoir, ensuring a physically complete and scalable representation of an open quantum system with built-in non-Markovian feedback and tunable amplification.

**Organization**

Section II derives the self-energy from first principles. Section III contrasts our architectural approach with reconstructive and geometric methods, emphasizing complementarity rather than replacement. Section IV presents numerical validation of engineered amplification. Section V outlines a synergistic workflow enabled by this architectural core.

**II. From Bath Correlations to Self-Energy**

The structured-bath model presented here is intentionally finite and simplified. This is not a concession but a mechanism. As demonstrated by our recent work[9], [10], coherent memory arises not from statistical vastness but from architectural constraint. By limiting coupling and discretizing the bath, we enable localized coherence traps that retain and recycle energy. This simplification is essential. In fact, it preserves the physical identity of dissipation and feedback, ensuring that the self-energy remains traceable to real couplings. In contrast to reconstructive approaches that infer coherence from fitted kernels or virtual chains, our model maintains quantum reciprocity by design. It is not the size of the bath that matters; it is the structural fidelity of its connection.

We begin with the linear system–bath Hamiltonian

$$H = H_S + \sum_k \omega_k b_k^\dagger b_k + \sum_k (g_k L^\dagger b_k + g_k^* b_k^\dagger L), \quad (1)$$

where $L$ is a system operator and $b_k$ are bath modes. Defining the collective bath operator $B = \sum_k g_k b_k$, the bath correlation function is

$$C(t) = \langle B(t) B^\dagger(0) \rangle. \quad (2)$$

In the continuum limit, with spectral density

$$J(\omega) = \sum_k |g_k|^2 \delta(\omega - \omega_k) \quad (3)$$

one obtains

$$C(t) = \int_0^\infty d\omega\, J(\omega)\, e^{-i\omega t}. \quad (4)$$

At finite temperature, this generalizes to:

$$C(t) = \int d\omega\, J(\omega)[(n(\omega) + 1)e^{-i\omega t} + n(\omega)e^{+i\omega t}], \quad n(\omega) = \frac{1}{e^{\hbar\omega/k_B T} - 1}. \quad (5\text{-}6)$$

The bath correlation $C(t)$ encodes both fluctuation and dissipation, and its frequency-domain representation defines the self-energy $\Sigma(\omega)$. The above formulation preserves detailed balance by explicitly including both emission $(n(\omega) + 1)$ and absorption $n(\omega)$ terms, weighted by the Bose-Einstein distribution. These terms reflect the thermal symmetry of the bath and ensure that the correlation function $C(t)$ satisfies the Kubo-Martin-Schwinger condition[11], [12]. Unlike reconstructive methods where detailed balance is imposed through fitting, this structure emerges directly from the finite Hamiltonian, maintaining physical traceability and architectural reciprocity. This self-energy is not imposed by fitting or reconstruction. It is derived directly from the finite Hamiltonian, preserving the physical origin of both gain and loss.

Although the continuum limit is invoked to express the spectral density and correlation function in integral form, this does not compromise the first-principles foundation of the formalism. The spectral density $J(\omega)$ itself is derived from a finite, structured Hamiltonian with physically meaningful couplings and mode distributions. The continuum representation serves as a compact encoding of this structure, not a replacement for it. As such, the resulting correlation function $C(t)$ and self-energy $\Sigma(\omega)$ retain a direct link to the underlying architecture of the system and bath.

While this formalism is broadly accepted, the way we construct the self-energy determines whether we retain a physical connection to the system or lose it to abstraction [13], [14]. In this framework, the self-energy is not a post-processed artifact but a direct consequence of structured coupling. This preserves what might be called the 'GPS of coherence', linking gain and loss to real, engineered layers rather than to reconstructed kernels. The distinction is critical. It governs whether quantum reciprocity is upheld as a matter of first-principle architecture or delegated to indirect means.

In this study, we apply a structured bath construction to derive the self-energy directly from the finite Hamiltonian. The real and imaginary parts of $\Sigma(\omega)$ govern, respectively, level renormalization and linewidth broadening. This approach eliminates bath coordinates at the Hamiltonian level without invoking external dispersion relations or analytic continuation. The complete derivations for both configurations are provided in Appendix 1.

For a structured finite environment, eliminating bath coordinates yields the architectural self-energy. For the two-layer chain:

$$S \overset{J_{SL_1}}{\leftrightarrow} L_1 \overset{J_{L_{12}}}{\leftrightarrow} L_2,$$

with weak losses to Markovian backgrounds $(\gamma_1, \gamma_2)$:

$$\Sigma_S(\omega) = \frac{J_{SL_1}^2}{\omega - \omega_{L_1} - \frac{J_{L_{12}}^2}{\omega - \omega_{L_2} + i\gamma_2} + i\gamma_1} \quad (7)$$

The nested denominator ties dissipation and feedback to the same microscopic channels $J_{SL_1}, J_{L_{12}}$, producing coupled $\text{Re}\Sigma(\omega)$ and $i\text{Im}\Sigma(\omega)$ that yield the breathing dynamics of quantum amplifiers. A worked derivation is provided in Appendix 1 (Case 1).

For an extended six-node network (one system, two layer-1 resonators (labeled $B_1$ and $B_2$), three layer-2 amplifier nodes (labeled $B_3, B_4$ and $B_5$),

$$\Sigma_S \omega = J_{SB}^T \mathbf{M}^{-1} J_{SB} \quad (8)$$

where $J_{SB}$ defines the bath-system coupling and $\mathbf{M}$ captures the bath influence and response. This shows that topology, not statistical size, governs coherence circulation. The exact matrix derivation appears in Appendix 1 (Case 2).

### III. The Loss of Reciprocity in Reconstructive Methods

In the previous section, we derived the self-energy Σ(ω) directly from the structured-bath Hamiltonian, where each coupling retains spatial identity and architectural meaning. We now examine the same Σ(ω) through the lens of widely used reconstructive methods—namely HEOM, TEMPO, and, to a limited extent, EPR—to understand how architectural features may be obscured when coherence is inferred rather than derived.

Solvers such as HEOM and TEMPO[6], [7], [8], offer powerful reconstructions of bath influence. and have been instrumental in modeling open quantum systems. However, these approaches typically begin from spectral data and build outward, rather than encoding the physical origin of dissipation and feedback. This distinction is subtle but significant. When the self-energy is assembled from fitted kernels or virtual chains, coherence often emerges as a statistical effect rather than a structural feature. EPR, while not dynamic, similarly reconstructs mode identity from static field distributions, without capturing the imaginary component of Σ(ω) or enforcing coupling symmetry. In contrast, the structured-bath Hamiltonian derives both gain and loss from the same microscopic couplings, preserving the architectural identity of the interface and allowing coherence to circulate. This is not a critique of numerical sophistication, but a reminder that physical agency must be retained if reciprocity is to be designed rather than inferred.

The principle of quantum reciprocity is tied to a Hamiltonian description where dissipation and feedback share a common microscopic origin. Leading non-Markovian techniques—the

Hierarchical Equations of Motion (HEOM)[6], [7] and tensor-network methods (e.g., TEMPO)[8]—while powerful, are reconstructive: they build a numerical image of the bath's influence on a posteriori and lose the physical interface that enables coherent feedback.
In HEOM, the bath correlation is approximated by a sum of exponentials,

$$C(t) \approx \sum_j c_j e^{-\gamma_j t} \quad (9)$$

$$\Sigma_{\text{HEOM}}(\omega) = \sum_j \frac{c_j}{\gamma_j - i\omega} \quad (10)$$

See Appendix 2 for derivations. Each term is a unidirectional decay channel. The resulting $\Sigma_{\text{HEOM}}$ is a sum of simple poles whose reactive and dissipative parts are linked mathematically but not by a shared microscopic coupling; the connection to a tunable architectural element is severed.

Similarly, tensor-network (TN/TEMPO) methods map the environment to a one-dimensional virtual chain,

$$H_{\text{chain}} = H_S + \sum_{n \geq 0} \epsilon_n b_n^\dagger b_n + \sum_{n \geq 0} (t_n b_n^\dagger b_{n+1} + h.c.) + \lambda(L^\dagger b_0 + b_0^\dagger L), \quad (11)$$

yielding a directional continued-fraction self-energy

$$\Sigma_{\text{TN}}(\omega) = \cfrac{\lambda^2}{\omega - \epsilon_0 - \cfrac{t_0^2}{\omega - \epsilon_1 - \cfrac{t_1^2}{\omega - \epsilon_2 - \cdots}}} \quad (12)$$

Also see Appendix 2 for derivations. Information flows forward along a virtual chain rather than through a physical interface capable of returning. In both cases, the self-energy is assembled from effective dissipation channels. The fundamental identity of the structured-bath Hamiltonian—that the same couplings $J_{SL_1}, J_{L_{12}}$ govern both Re Σ(reactive) and Im Σ(dissipative)—is effectively masked. This is not merely a technical omission. It represents a conceptual blind spot. Without a shared microscopic origin, coherence cannot circulate. It can only decay.

Beyond the loss of reciprocity, these reconstructive methods also obscure spatial identity, suppress topological control, and limit predictive design. Because virtual chains and fitted kernels lack physical coordinates, they cannot track where coherence resides or how it flows. This makes it difficult to engineer feedback paths or optimize amplifier geometry. Moreover, the absence of explicit couplings prevents parametric tuning, which is essential for gain control and mode selection. These limitations do not reflect numerical weakness, but architectural incompleteness. Without a structured interface, coherence remains a statistical outcome rather than a controllable feature.

Along this line of thought, it is also essential to discuss our approach in the context of the Energy Participation Ratio (EPR) framework [15], [16], which treats superconducting circuits as closed and lossless systems. In EPR, the electromagnetic modes of the circuit are first obtained from the classical field solutions, and the portion of energy stored in each element $i$ for mode $m$ is expressed through its participation factor

$$p_{i,m} = \frac{U_{i,m}}{U_{\text{tot},m}}, \quad (13)$$

where $U_{i,m}$ is the energy localized in element $i$ and $U_{\text{tot},m}$ is the total energy of mode $m$. The quantized Hamiltonian is then written as

$$H_{\text{EPR}} = \sum_m \hbar\omega_m a_m^\dagger a_m + \sum_{i,m,n} E_i\, p_{i,m} p_{i,n} (a_m + a_m^\dagger)(a_n + a_n^\dagger), \quad (14)$$

so that the frequency correction of mode $m$ becomes

$$\Delta\omega_m = \sum_i E_i p_{i,m}^2. \quad (15)$$

This correction can be regarded as the real part of an effective self-energy,

$$\boxed{\Sigma_{\text{EPR}}(\omega) \approx \Delta\omega_m} \quad (16)$$

which encapsulates the static renormalization of the mode frequency due to the electromagnetic loading of the circuit elements. See Appendix 3 for the derivations. Because the environment is not quantized, there is no corresponding imaginary term to describe feedback or dissipation. The EPR formalism therefore captures only static field participation and detuning; it provides a geometric map of how energy is distributed among elements, rather than a dynamical account of energy exchange or memory.

Our structured bath framework extends this picture by reintroducing the environment as an explicit quantum network. The full complex self-energy, as shown in Eq. 7, emerges directly from the finite Hamiltonian, where the real and imaginary parts share the same microscopic couplings that govern both frequency renormalization and dissipation. This converts the static detunings of EPR into active channels of reciprocity. Whereas EPR fixes the participation of each element once and for all, the structured bath allows those participations to evolve, preserving the spatial and energetic identity of each node and providing a continuous, measurable record of coherence flow across the amplifier architecture.

As summarized in Table 1, the structured-bath formulation preserves reciprocity by linking dissipation and feedback to the same microscopic couplings within a finite, quantized architecture. In contrast, HEOM and TN/TEMPO reconstruct the bath influence statistically, while the EPR framework captures only static field participation without dynamic memory. Together these distinctions clarify why coherent amplification and reciprocal energy exchange require an explicit quantum architecture rather than a reconstructed or geometric approximation.

# Table 1 — Conceptual Comparison of Open-System Frameworks

| Framework | Type and Scope | Construction of $\Sigma\omega$ | Microscopic Link to System | Treatment of Dissipation and Feedback | Physical Interpretation |
|---|---|---|---|---|---|
| **Structured Bath (this work)** | Finite quantized architecture with explicit near-field layers | Derived directly from finite Hamiltonian | Shared microscopic couplings $J_{SL_1}, J_{L_{12}}$ generate both $\text{Re}\Sigma(\omega)$ and $i\text{Im}\Sigma(\omega)$ | Dissipation and feedback emerge from same channels, producing coherent amplification and reciprocity | Dynamical, quantized near-field environment retaining spatial and energetic identity |
| HEOM | Hierarchical expansion of reduced dynamics | $\Sigma_{HEOM}(\omega) = \sum_j \frac{c_j}{\gamma_j - i\omega}$ from exponential decomposition of C(t) (Eqs. 9–10) | Correlations reconstructed from fitted decay channels; no explicit bath nodes | Purely unidirectional memory terms; feedback absent | Numerically exact but reconstructive; memory captured statistically |
| TN / TEMPO | Tensor-network chain mapping of environment | Continued-fraction form $\Sigma_{TN}(\omega) = \dfrac{\lambda^2}{\omega - \epsilon_0 - \dfrac{t_0^2}{\omega - \epsilon_1 - \cdots}}$ | Virtual sites replace physical couplings | Directional information flow; no reciprocity | Efficient numerical encoding of non-Markovian tails |
| EPR | Closed, lossless circuit model based on classical field modes | $\Sigma_{EPR}(\omega) \approx \Delta\omega_m = \sum_i E_i p_{i,m}^2$ (Eqs. 15-16) | Static participation factors link elements to modes | Only real part of $\Sigma$ captured; no feedback or dissipation | Geometric map of field participation; static detuning, no dynamic |

The following section presents numerical validation of this architectural principle, demonstrating how engineered coupling within the structured bath reproduces the gain and coherence flow predicted by the theoretical framework**.**

## IV. Hamiltonian Reciprocity and Engineered Amplification

We further tested the architectural model with numerical simulations of the six-qubit structured-bath Hamiltonian (derivations in Appendix 1). The finite near-field bath is modeled explicitly, while boundary nodes couple weakly to a featureless Markovian environment. By tuning on-site energies and interlayer couplings, which is implemented as a parametric pump, we steer the device across operating regimes.

Key outcomes:

<u>Reciprocity in action.</u> The same architectural couplings $(J_{SL_1}, J_{L_{12}})$ jointly set Re $\Sigma$(frequency renormalization) and Im $\Sigma$(loss/gain) of the central system.

Engineered gain. We realize a controllable transition from passive dissipation to coherent amplification, quantified by the system Green's function

$$G_S(\omega) = [\omega - \omega_S - \Sigma_S(\omega)]^{-1} \quad (20)$$

Design maps. Pump sweeps and interlayer-coupling ratios generate gain/phase landscapes that visualize energy circulation and guide device optimization.

These simulated results confirm that architectural control of the near field governs the operational regime. Practically, optimization amounts to shaping the spectral response

$$\mathcal{G}(\omega; J_{L_1L_2}, P) = |G_S(\omega; \Sigma(\omega; J_{L_1L_2}, P))|^2 \quad (21)$$

where the self-energy $\Sigma(\omega; J_{L_1L_2}, P)$ carries the parametric dressing induced by the pump $P$ and interlayer coupling $J_{L_1L_2}$. By controlling three knobs—$J_{L_1L_2}$, pump strength $P$ (or $g_p$), and probe frequency $\omega$—we sculpt the poles of $G_S(\omega)$ and move the amplifier between under-coupled, critically-coupled, and over-coupled (gain) regimes at will.

Table 2 below summarizes three control parameters that define the amplifier's operational regime. The interlayer coupling $J_{L_1L_2}$ governs how energy flows between the storage layer ($L_1$) and the output layer ($L_2$), directly shaping the spectral poles of the system Green's function $G_S(\omega)$. The pump strength $g_p$ or power $P$ introduces nonlinear dressing into the bath, modifying the self-energy $\Sigma(\omega)$ and enabling transitions from passive to gain regimes. The probe frequency $\omega$ selects which dressed modes are interrogated, controlling detuning and phase matching. Together, these parameters form a minimal yet complete control set for navigating the amplifier's behavior—from quiet hold to bright amplification, from transparency to structured coherence.

**Table 2: Architectural Control Parameters: Mapping Physical Action to Spectral Response**

| Control | Physical meaning | Mathematical role | Experimental handle |
|---|---|---|---|
| $J_{L_1L_2}$ | Exchange between storage ($L_1$) and output ($L_2$) | Sets hybridization; shifts/merges poles of $G_S$ | Flux-tunable coupler or capacitive spacing |
| Pump ($g_p, P$) | Parametric drive enabling nonlinear conversion | Dresses $\Sigma$; sets gain, threshold, bifurcation | Microwave pump amplitude/power |
| $\omega$ | Probe frequency relative to dressed modes | Sets detuning and phase matching | Signal generator / LO or flux bias |

The numerical spectra that follow are all generated from the same six-node structured-bath Hamiltonian of Equation (8). The computational script allows the architectural parameters $J_{L_1L_2}, J_{L_2,\text{int}}, J_{SB}, \gamma_1, \gamma_2$, and the pump $P$ to be varied continuously, producing both the transparent baseline and the fully structured response as limiting cases of one model. Figure 1 corresponds to the uncoupled limit, where $J_{L_1L_2}$ and $J_{L_2}$ are nearly zero and the pump is inactive, yielding the Markovian, memory-free regime. Increasing these same couplings and activating the pump generates the structured-amplifier spectra discussed in Figure 2. Each color map shows the computed intensity $|G_{B_3 \leftarrow S}(\omega)|^2$ for the same underlying Hamiltonian. The color scale represents the power transferred from the system (S) to the amplifier output (B₃), while the bright ridge highlights the dominant hybridized mode that emerges from the coupled system–bath dynamics.

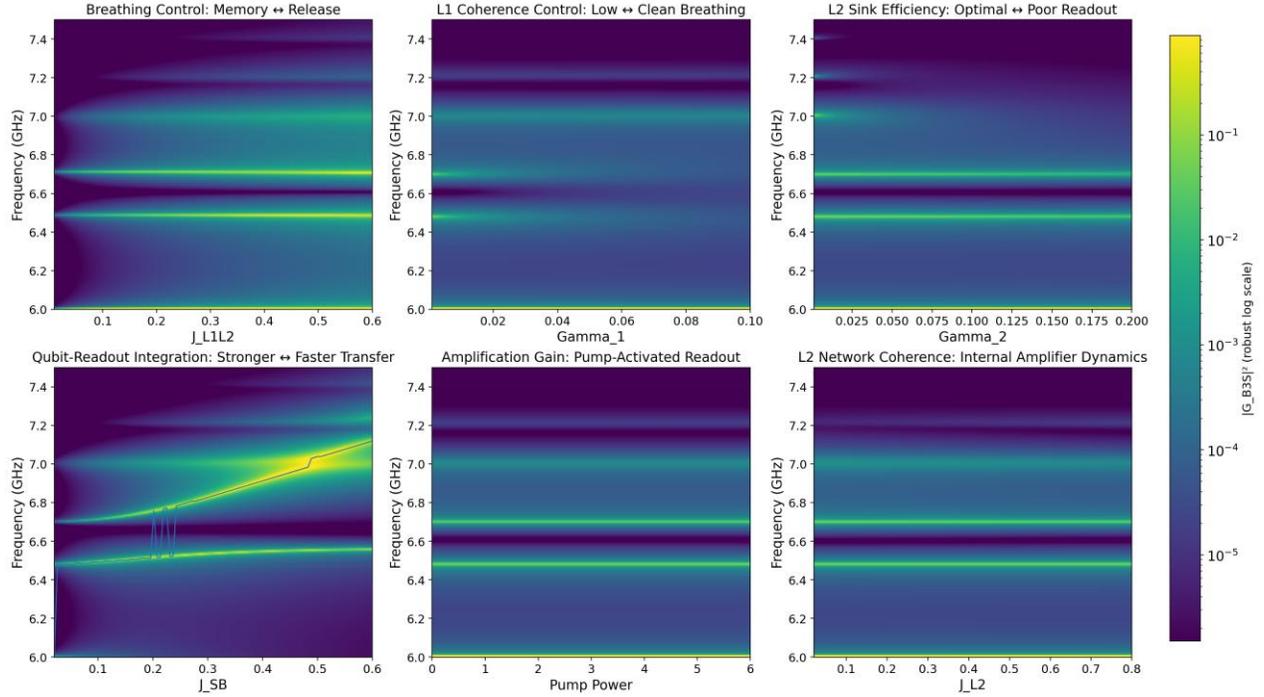

Figure 1. Isolated-System Baseline: Transparent Amplifier Without Internal Coupling

**Computational Workflow**
The complete computational workflow used to generate the spectra is available as open source in the accompanying GitHub repository. The script implements the full six-node structured-bath Hamiltonian and performs parameter sweeps over the architectural controls $J_{L_1L_2}$, $J_{L_2,\text{int}}$, $J_{SB}$, $\gamma_1$, $\gamma_2$, and the parametric drive P. For each sweep, the complex self-energy and gain function $|G_{B_3 \leftrightarrow S}(\omega)|^2$ are computed and stored as two-dimensional arrays, together with the ridge line identifying the maximum spectral response. In all parameter sweeps except the pump scan, the drive is fixed at P = 0.0, representing the passive baseline. During the pump scan, P is varied from 0 to 6 to capture the onset of active response. Extended parameter combinations and ridge maps are provided in Appendix 4.

**Figure 1 – Transparent, Markovian Baseline**
Figure 1 represents the transparent, memory-free regime. The system–bath couplings are weak ($J_{SB_1,SB_2} = 0.005, 0.01$); the first-layer bridge is minimal ($J_{B_1B_2} = 0.05$); and the second-layer triangle is nearly inactive ($J_{L_1L_2} \approx 0.05$–$0.06$, $J_{L_2,\text{int}} = 0.01$). Dissipation is small (γ₁ = γ₂ = 10⁻³) and no pump field is applied. Under these conditions the architecture behaves as a transparent, memory-free amplifier. Energy emitted from the system flows directly outward through $L_1$ and $L_2$ without feedback or re-absorption. The resulting spectra display bright, nearly horizontal ridges centered around 6.4–6.6 GHz—the resonant response of the qubit–$L_1$ pair—with weaker shoulders near 7.0 GHz from residual coupling to $L_2$. Because inter-layer correlations are negligible, there are no dark bands or interference fringes: each bath node radiates independently, and the overall output is the linear superposition of these one-pass channels. The ridge line in each panel traces the frequency of maximum gain,

$$\omega_{\text{ridge}} = \arg \max_{\omega} \mid G_{B_3S}(\omega) \mid^2 \quad (22)$$

and appears flat across all sweeps, indicating that the effective self-energy $\Sigma_S(\omega)$ is nearly constant. This flat, featureless profile defines the Markovian baseline of the amplifier—an emission regime without stored coherence or reciprocal feedback. Only when the couplings $J_{L_1L_2}$ or $J_{L_2,\text{int}}$ are strengthened and the pump is activated (as in later configurations) do diagonal ridges, anti-crossings, and gain tongues emerge, signaling the onset of hybridization and coherent amplification.

The corresponding heatmap (e.g. $B_3$ vs $J_{L_1L_2}$) shows a smooth, continuous ridge with no deep cancellation zones. This confirms that every photon emitted from either $B_1$ or $B_2$ reaches the output without interference. The result is full constructive addition of amplitudes and a clean, nearly linear spectrum. This is the Markovian baseline of the amplifier: energy exits immediately, leaving no trace of memory or recycling.

As coupling parameters are swept, this isolated regime transitions into a structured amplifier. Increasing $J_{L_1L_2}$ transforms flat modes into diagonal streaks—evidence of level repulsion and energy flow between $L_1$ and $L_2$. Adjusting $\gamma_1$ or $\gamma_2$ modulates brightness, controlling how efficiently energy is passed or absorbed. Strengthening $J_{SB}$ splits the system–bath modes, while activating the pump bends resonances into a bright anti-crossing that marks the gain maximum. Finally, increasing $J_{L_2}$ internal interference fringes, signaling the emergence of internal memory and coherent recycling.

**Structured Regime and Amplification (Figure 2)**
Figure 2 presents the structured-amplifier regime obtained from the same six-node Hamiltonian used for Figure 1, but now with finite inter-layer couplings and an active parametric drive. The parameters are $J_{L_1L_2} = 0.05 - 0.30, J_{L_2,\text{int}} = 0.02 - 0.10 J_{SB1,SB2} = 0.005, 0.010, \gamma_1 = \gamma_2 = 10^{-3}, P \in [0,6]$. These conditions reveal how coupling geometry and pump strength together shape coherence storage, feedback, and gain.

Even before activating the pump field, the structured configuration exhibits a higher overall intensity scale than the transparent baseline of Figure 1. This apparent passive gain arises from constructive interference among multiple coupling paths. Finite $J_{L_1L_2}$ and $J_{L_2,\text{int}}$ increase the local density of states and concentrate energy flow toward the output node $B_3$, leading to a stronger $\mid G_{B_3S}(\omega) \mid^2$ response even at $P = 0$. The architecture itself—without drive—thus redistributes and amplifies spectral power through coherent hybridization.

Figure 2 reveals how structured coupling transforms the flat response of Figure 1 into a coherent interference landscape. Increasing $J_{L_1L_2}$ causes energy to circulate between the inner and outer bath layers, producing bright ridges and dark nodes that record periodic exchange. Low damping ($\gamma_1, \gamma_2$) sustains this oscillatory "breathing," while higher damping converts it into stable gain channels. The System–Bath Coupling panel shows that weak $J_{SB}$ isolates the system, yielding narrow resonances, whereas stronger coupling broadens and splits them, marking hybridization and faster energy transfer. The Pump Power panel demonstrates how the parametric drive activates the amplifier: a bright gain tongue appears near 6.6–6.8 GHz, sharpening and then saturating as P increases, yet remaining quantum-coherent. Finally, increasing $J_{L_2,\text{int}}$ introduces fine interference fringes—signatures of energy recirculation within the outer bath that sustain long-lived feedback loops. The structured bath thus converts dissipation into coherent recycling, bridging passive response and active amplification.

### Figure 3 – Pump-Driven Amplification and Spectral Compression

Figure 3 extends the structured regime of Figure 2 by increasing both inter-layer and intra-layer connectivity while preserving the same six-node Hamiltonian. The bridge within the first-layer network is reinforced ($J_{B_1B_2} = 0.4$), the inter-layer links remain strong ($J_{L_1L_2} = 0.45$), and the triangular feedback loop of the amplifier layer is slightly tightened ($J_{L_2} = 0.15$). The system–bath channels are set to ($J_{B_1B_2} = 0.30, 0.26$), with low dissipation γ₁ = 10⁻³ and γ₂ = 2 × 10⁻². The pump amplitude P is swept from 0 to 6, acting through an effective coupling $J_{34}^{eff} = J_{34} + gP$ with $g$ = 0.2.

Under these conditions, the amplifier transitions from a hybrid breathing regime to a coherent, gain-dominated state. Even at zero drive, the strong $J_{B_1B_2}$ bridge compresses the hybridized spectrum into a narrow ridge centered on the output node B₃, establishing a preferred channel for coherent transfer. As the pump increases, the competition between the L₁ bridge and L₂ feedback loop forming a bright triangular-like *gain tongue* (see arrow) centered near 6.6–6.8 GHz and $P \approx 2 - 3$. This region defines the onset of parametric amplification where coherence is reinforced by internal feedback rather than dissipated.

This simulated "gain tongue" corresponds directly to the experimentally observed pump–frequency phase diagram used to calibrate Josephson parametric amplifiers. In experiments, the tongue marks the stability window between the onset of parametric gain and the threshold of bifurcation. Its triangular shape, bounded by a critical pump amplitude, is routinely used to extract the effective nonlinear coupling and dissipation balance in devices such as flux-driven Josephson amplifiers and traveling-wave parametric amplifiers[17].

To our knowledge, this is the first demonstration in which the experimentally observed gain tongue of a Josephson-like quantum amplifier emerges directly from a first-principles Hamiltonian. The structured-bath formulation derives the amplification window from the same microscopic couplings that also generate dissipation, without relying on phenomenological stability equations. This approach unifies coherence, feedback, and loss within a single quantized architecture, providing a microscopic origin for the gain–stability diagram that has long defined experimental quantum amplifiers.

Overall, Figures 2 and 3 map a continuous evolution of the structured bath—from passive redistribution to pump-assisted amplification—showing that the same microscopic couplings responsible for dissipation in the weak-coupling limit can, through reciprocity, generate controlled quantum gain. The *gain-tongue* region observed here corresponds directly to the experimentally established pump–frequency stability map of Josephson-like amplifiers and is discussed further in Appendix 4, where its geometric origin is analyzed as a hallmark of architectural feedback.

This transition—from passive emission to structured amplification—is governed by architectural suppression and release, not by emergent complexity. The amplifier's behavior arises from the interplay of geometry, inter-layer coupling, and parametric drive. As the pump engages, the inter-layer links $J_{L_1L_2}$ act as programmable gates: small values suppress exchange and store coherence; intermediate values enable breathing and feedback; large values convert stored coherence into gain. This controlled modulation, rather than brute-force drive, defines the amplifier's operational fingerprint—the boundary between transparency and structure, between emission and interference.

The computational workflow that produces these spectra is based on the open structured-bath Hamiltonian. It performs parameter sweeps over the architectural controls $J_{L_1L_2}, J_{L_2}, J_{SB}, \gamma_1, \gamma_2,$ and $P$, and exports the resulting gain and ridge data as two-dimensional arrays corresponding to the heat-map panels in Figures 1–3 and the Appendices. In all parameter sweeps except the pump scan, the parametric drive is fixed at $P = 0$, representing the amplifier's steady operating condition; during the pump sweep, $P$ is varied continuously from 0 to 6 to capture the full gain transition.

**V. Outlook**
The simulated spectra above demonstrate how architectural suppression and release, coherent storage followed by directed amplification, emerge directly from the structured bath Hamiltonian. The analysis therefore offers a self-consistent framework in which coherence, feedback, and dissipation arise from the same microscopic origin. When the near-field environment is treated as a finite structured and quantized network, the usual boundary between system and bath becomes an internal feature of the coupling topology rather than an imposed constraint. The near field consequently acts as a physically resolved mediator whose reciprocity is defined by geometry and interaction strength, providing a direct link between microscopic design and macroscopic response.

Extending this description beyond the explicitly simulated region will require connecting the quantized near field dynamics to the statistical continuum that characterizes realistic device operation. In this context, the structured bath serves as a first principles anchor for multiscale modeling, where learned leakage parameters and hierarchical solvers can translate the finite architectural layer into predictive and experimentally accessible quantities. Within this framework, the structured-bath formulation defines the near field as the region where coherence, feedback, and dissipation emerge from the same microscopic Hamiltonian. Treating the environment as a finite quantized network makes the system–bath boundary an internal feature of the coupling topology. The resonator–amplifier pair ($L_1$, $L_2$) forms the architectural core that governs how quantum energy is stored, delayed, and released through geometry.

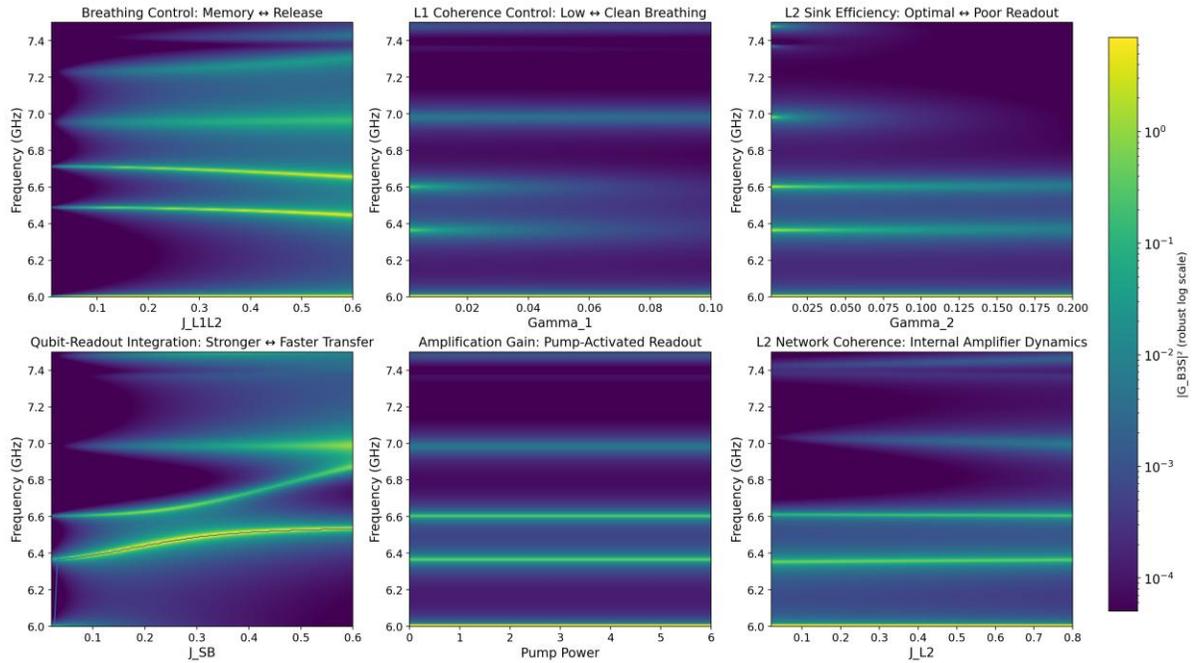

**Figure 2 – Structured Regime and Coherent Breathing.**
With finite inter-layer coupling ($J_{L_1L_2} \approx 0.05 - 0.3$) and low damping, alternating bright and dark ridges appear, showing coherent energy exchange between bath layers. The network remains passive, with no amplification.

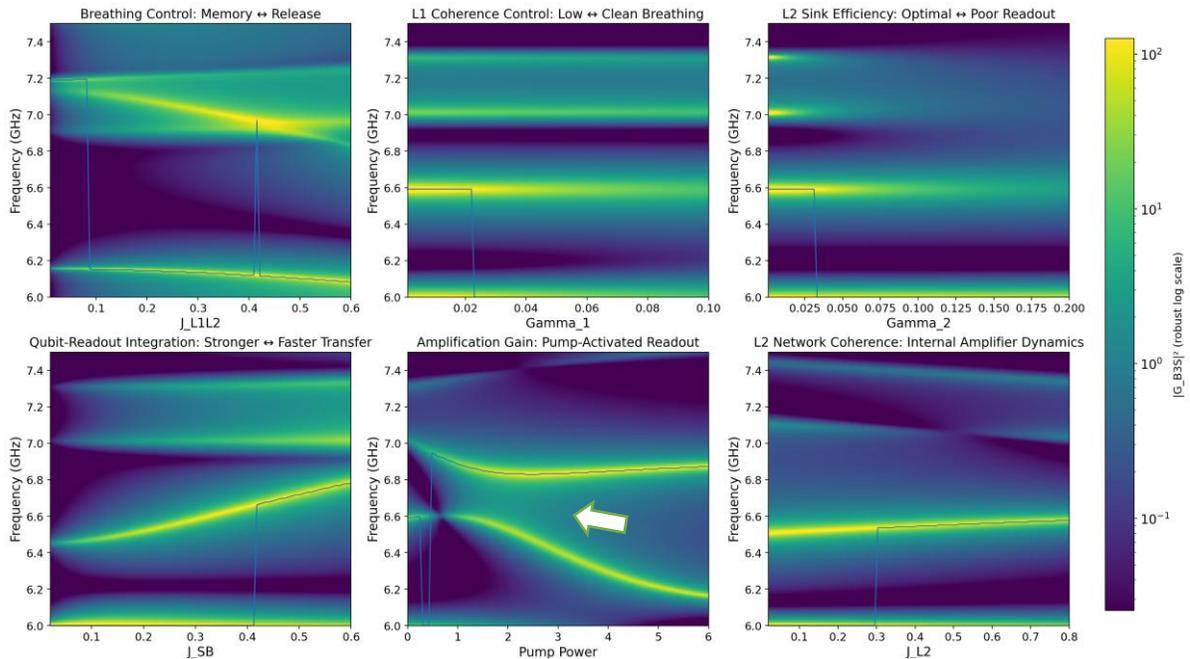

**Figure 3 – Pump-Driven Amplification and Gain Tongue Formation.**
At high coupling ($J_{B_1B_2} = 0.4, J_{L_1L_2} = 0.45, J_{L_2} = 0.15$), a bright triangular-like gain tongue (arrow) appears near 6.6–6.8 GHz for $P \approx 2-3$, marking the onset of parametric amplification. For larger $P$ in this range, the ridge shape remains essentially unchanged within the plotting resolution.

This near-field layer constitutes the first level of the multiscale architecture: its Hamiltonian, parameterized by explicit couplings $J_{ij}$ and local frequencies $\omega_i$, captures the dominant non-Markovian interactions and intrinsic reciprocity at the interface. It provides the self-energy that mediates coherent exchange between the system and its dissipative surroundings, reproducing the experimentally observed dynamics of gain and delay.

At the second level, this microscopic architecture supports a machine-learning bridge that connects simulation and experiment. The gain and phase-response maps generated from the structured model can be used to infer effective leakage parameters $\gamma_i$ that characterize how the engineered near field couples to the far-field environment. These learned quantities translate the geometric design of the interface into measurable dissipation channels, allowing the structured bath to function as a generator of physically meaningful input for larger-scale solvers.

At the third level, formal hierarchical solvers such as HEOM or TN/TEMPO can propagate the remaining weakly coupled continuum once the near-field channels have been parameterized. In this hierarchy, the structured bath defines the quantized gateway of reciprocity, while HEOM and TN continue its dynamics statistically. Their kernels and tensor decompositions maintain numerical completeness without sacrificing the physical traceability established at the microscopic layer.

At the static limit, this workflow converges naturally with the Energy Participation Ratio (EPR) framework used in circuit quantization, where stored electromagnetic energy defines fixed participation ratios among circuit elements. The structured-bath formulation generalizes this concept into the temporal domain: participation factors evolve dynamically, encoding the reversible exchange between coherent and dissipative channels. The near-field geometry thus becomes a dynamic rather than static determinant of quantum energy distribution.

This multiscale integration transforms environmental modeling into a reproducible design principle. The structured near field provides the quantized kernel of coherence, the machine-learning bridge converts its spectral fingerprints into measurable parameters, and hierarchical solvers extend its reach into the continuum limit. Together they establish a continuous, first-principles pathway from microscopic reciprocity to macroscopic control, enabling quantum devices in which coherence, memory, and noise are not residual effects but deliberately engineered properties.

**Summary**

The resulting formulation transforms quantum noise from a boundary condition into a tunable design parameter. The structured bath defines a quantized interface through which coherence, feedback, and amplification can be shaped by geometry. By unifying the near field and far field descriptions under one Hamiltonian framework, this approach provides a physically traceable and experimentally adaptable foundation for quantum device design, where loss, memory, and reciprocity are co engineered manifestations of the same microscopic architecture.

**Code and Data Availability**

All simulation codes and data supporting this study are available at the GitHub repository Quantum-Amplification (https://github.com/sakidja/Quantum-Amplification). This repository extends the earlier quantum_bath project (https://github.com/sakidja/quantum_bath) developed for the foundational Structured Quantum Baths framework.

**Acknowledgement**

This research used resources of the National Energy Research Scientific Computing Center (NERSC) through the QIS@Perlmutter program. The support enabled simulations advancing the study of open quantum systems and non-Markovian dynamics. This study also benefited from engaging discussions and assistance in code development with large language models (LLMs).

**References**

[1] M. H. Devoret, J. M. Martinis, and J. Clarke, "Measurements of Macroscopic Quantum Tunneling out of the Zero-Voltage State of a Current-Biased Josephson Junction," *Phys. Rev. Lett.*, vol. 55, no. 18, pp. 1908–1911, Oct. 1985, doi: 10.1103/PhysRevLett.55.1908.

[2] W. Zhixin, "From artificial atoms to quantum information machines: Inside the 2025 Nobel Prize in physics," The Conversation. [Online]. Available: https://theconversation.com/from-artificial-atoms-to-quantum-information-machines-inside-the-2025-nobel-prize-in-physics-266976

[3] W. H. Zurek, "Decoherence, einselection, and the quantum origins of the classical," *Rev. Mod. Phys.*, vol. 75, no. 3, pp. 715–775, May 2003, doi: 10.1103/RevModPhys.75.715.

[4] Y. Zeng, J. Stenarson, P. Sobis, and J. Grahn, "Pulsed HEMT LNA Operation for Qubit Readout," *IEEE Transactions on Microwave Theory and Techniques*, vol. 73, no. 9, pp. 6539–6553, Sept. 2025, doi: 10.1109/TMTT.2025.3556982.

[5] SciTechDaily Staff, "New Quantum Amplifier Uses 90% Less Power – Without Sacrificing Performance." [Online]. Available: https://scitechdaily.com/new-quantum-amplifier-uses-90-less-power-without-sacrificing-performance/

[6] Y. Tanimura and R. Kubo, "Time Evolution of a Quantum System in Contact with a Nearly Gaussian-Markoffian Noise Bath," *J. Phys. Soc. Jpn.*, vol. 58, no. 1, pp. 101–114, Jan. 1989, doi: 10.1143/JPSJ.58.101.

[7] Y. Tanimura, "Numerically 'exact' approach to open quantum dynamics: The hierarchical equations of motion (HEOM)," *The Journal of Chemical Physics*, vol. 153, no. 2, p. 020901, July 2020, doi: 10.1063/5.0011599.

[8] A. Strathearn, P. Kirton, D. Kilda, J. Keeling, and B. W. Lovett, "Efficient non-Markovian quantum dynamics using time-evolving matrix product operators," *Nature Communications*, vol. 9, no. 1, p. 3322, Aug. 2018, doi: 10.1038/s41467-018-05617-3.

[9] R. Sakidja, "Structured Quantum Baths with Memory: A QuTiP Framework for Spectral Diagnostics and Machine Learning Inference," *arXiv preprint arXiv:2508.17514*, 2025.

[10] R. Sakidja, *Quantum Bath Simulation (QuTiP)*. (Aug. 2025). Zenodo. doi: 10.5281/zenodo.16938479.

[11] R. Kubo, "Statistical-Mechanical Theory of Irreversible Processes. I. General Theory and Simple Applications to Magnetic and Conduction Problems," *J. Phys. Soc. Jpn.*, vol. 12, no. 6, pp. 570–586, June 1957, doi: 10.1143/JPSJ.12.570.

[12] P. C. Martin and J. Schwinger, "Theory of Many-Particle Systems. I," *Phys. Rev.*, vol. 115, no. 6, pp. 1342–1373, Sept. 1959, doi: 10.1103/PhysRev.115.1342.

[13] H. P. Breuer and F. Petruccione, *The Theory of Open Quantum Systems*. Oxford University Press, 2002. [Online]. Available: https://books.google.com/books?id=0Yx5VzaMYm8C

[14] A. J. Leggett, S. Chakravarty, A. T. Dorsey, M. P. A. Fisher, A. Garg, and W. Zwerger, "Dynamics of the dissipative two-state system," *Rev. Mod. Phys.*, vol. 59, no. 1, pp. 1–85, Jan. 1987, doi: 10.1103/RevModPhys.59.1.

[15] Z. K. Minev, Z. Leghtas, S. O. Mundhada, L. Christakis, I. M. Pop, and M. H. Devoret, "Energy-participation quantization of Josephson circuits," *npj Quantum Information*, vol. 7, no. 1, p. 131, Aug. 2021, doi: 10.1038/s41534-021-00461-8.

[16] M. Reagor *et al.*, "Quantum memory with millisecond coherence in circuit QED," *Phys. Rev. B*, vol. 94, no. 1, p. 014506, July 2016, doi: 10.1103/PhysRevB.94.014506.

[17] Ananda Roy and Michel Devoret, "Introduction to parametric amplification of quantum signals with Josephson circuits," *Comptes Rendus. Physique*, vol. 17, no. 7, pp. 740–755, 2016, doi: 10.1016/j.crhy.2016.07.012.

# Appendix 1 — Derivation of Self-Energy for Structured Baths

This appendix contains the detailed mathematical derivations for the simple chain and the six-qubit Hamiltonian, including Green-function equations, Schur complements, and the matrix identity with all intermediate steps.

CASE1: A SIMPLE CHAIN 3 QUBIT STRUCTURED BATH HAMILTONIAN

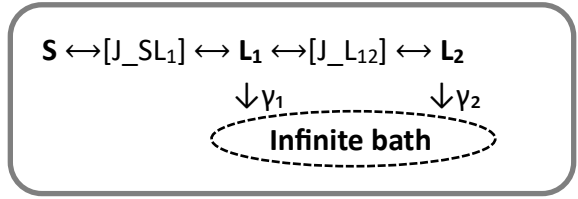

Step 1: Green's function equations:

$$(\omega - \omega_S)G_{SS} = 1 + J_{SL_1}G_{L_1 S} \quad (A1)$$

$$(\omega - \omega_1 + i\gamma_1)G_{L_1 S} = J_{SL_1}G_{SS} + J_{L_{12}}G_{L_2 S} \quad (A2)$$

$$(\omega - \omega_2 + i\gamma_2)G_{L_2 S} = J_{L_{12}}G_{L_1 S} \quad (A3)$$

Step 2: Substitute $G_{L_2 S}$ with $G_{L_1 S}$ into A2 equation

From (A3):

$$G_{L_2 S} = \frac{J_{L_{12}}}{(\omega - \omega_2 + i\gamma_2)} G_{L_1 S} \quad (A4)$$

Substituting into (A2):

$$(\omega - \omega_1 + i\gamma_1)G_{L_1 S} = J_{SL_1}G_{SS} + J_{L_{12}} \frac{J_{L_{12}}}{(\omega - \omega_2 + i\gamma_2)} G_{L_1 S} \quad (A5)$$

$$\left(\omega - \omega_1 + i\gamma_1 - \frac{J_{L_{12}}^2}{(\omega - \omega_2 + i\gamma_2)}\right) G_{L_1 S} = J_{SL_1}G_{SS} \quad (A6)$$

Hence:

$$G_{L_1 S} = \frac{J_{SL_1}}{\left(\omega - \omega_1 + i\gamma_1 - \frac{J_{L_{12}}^2}{(\omega - \omega_2 + i\gamma_2)}\right)} G_{SS} \quad (A7)$$

Step 3: Substitute $G_{L_1 S}$ into the first equation (A1)

$$(\omega - \omega_S)G_{SS} = 1 + J_{SL_1} \frac{J_{SL_1}}{\left(\omega - \omega_1 + i\gamma_1 - \frac{J_{L_{12}}^2}{(\omega - \omega_2 + i\gamma_2)}\right)} G_{SS} \quad (A8)$$

$$\left(\omega - \omega_S - \underbrace{\frac{J_{SL_1}^2}{\left(\omega - \omega_1 + i\gamma_1 - \frac{J_{L_{12}}^2}{(\omega - \omega_2 + i\gamma_2)}\right)}}_{\Sigma_S \omega}\right) G_{SS} = 1 \quad (A9)$$

Hence:

$$G_{SS} = \frac{1}{\omega - \omega_S - \Sigma_S \omega} \quad (A10)$$

with the self-energy:

$$\boxed{\Sigma_S \omega = \omega - \omega_1 + i\gamma_1 - \frac{J_{L_{12}}^2}{(\omega - \omega_2 + i\gamma_2)}} \quad (A11).$$

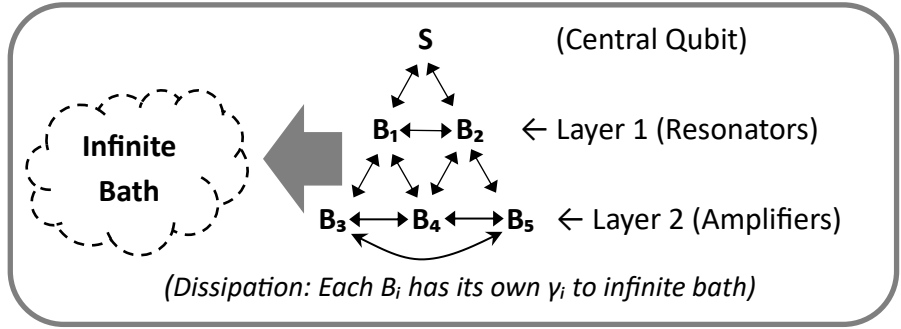

CASE2: 6 QUBIT STRUCTURED BATH HAMILTONIAN

S (Central Qubit)
$B_1 \leftrightarrow B_2$ ← Layer 1 (Resonators)
$B_3 \leftrightarrow B_4 \leftrightarrow B_5$ ← Layer 2 (Amplifiers)

(Dissipation: Each $B_i$ has its own $\gamma_i$ to infinite bath)

Modeling strategies:

Each bath site $B_i$ is coupled to a Markovian continuum with rate $\gamma_i$; we model this as onsite loss, so the resolvent diagonal is $\omega - \omega_i + i\gamma_i$. Thus $\gamma_i$ is the half-width at half-maximum (HWHM) of the Lorentzian response of $B_i$ and equals half the total energy-decay rate into the continuum. Equivalently, with a Lindblad operator $L_i = \sqrt{\kappa_i}\,\sigma^-_{B_i}$ one has a linewidth $\kappa_i$ and the same Green's-function diagonal ($\omega - \omega_i + i\kappa_i/2$); we simply write $\gamma_i \equiv \kappa_i/2$. The external environment contributes only to this small, memoryless decay. All non-Markovian structure is kept explicitly inside the finite bath graph (L₁/L₂), so the only memory relevant to our results is generated by the structured, quantized bath itself. This is precisely the mechanism we exploit for quantum-memory functionality: by engineering the internal couplings and mode detunings of the structured bath, we shape the system's self-energy to retain and release excitations on demand.

Beyond static design, the pump dynamically programs the structured bath *in situ*. In practice, the drive is applied to node B₃, which in a real amplifier corresponds to the flux- or voltage-biased element that sets the local resonance. The pump can shift the on-site frequency of B₃, $\omega_3 \to \omega_3 + \delta\omega_3$, through Stark-like tuning of its bias, and it can also modulate the couplings $J_{34} \to J_{34} + \delta J_{34}$ and $J_{35} \to J_{35} + \delta J_{35}$ by altering the effective impedances of the connecting junctions or resonator segments. The level-2 bath forms a triangular network linking B₃, B₄, and B₅ through $J_{34}$, $J_{45}$, and $J_{35}$; this geometry allows a single pump tone applied at B₃ to redistribute coherence across the entire triangle, thereby establishing the internal feedback loop that governs gain and phase stability.

System-Bath couplings
$$H_{SB} = J_{SB_1}\left(\sigma^+_S \sigma^-_{B_1} + \sigma^-_S \sigma^+_{B_1}\right) + J_{SB_2}\left(\sigma^+_S \sigma^-_{B_2} + \sigma^-_S \sigma^+_{B_2}\right) \quad (A12)$$

Layer 1 Couplings:
$$H_{L_1} = J_{B_1 B_2}\left(\sigma^+_{B_1} \sigma^-_{B_2} + \sigma^-_{B_1} \sigma^+_{B_2}\right) \quad (A13)$$

Layer 2 Couplings (pairwise-interconnected):
$$H_{L_2} = J_{B_3 B_4}\left(\sigma^+_{B_3}\sigma^-_{B_4} + \sigma^-_{B_3}\sigma^+_{B_4}\right) + J_{B_4 B_5}\left(\sigma^+_{B_4}\sigma^-_{B_5} + \sigma^-_{B_4}\sigma^+_{B_5}\right) + J_{B_3 B_5}\left(\sigma^+_{B_3}\sigma^-_{B_5} + \sigma^-_{B_3}\sigma^+_{B_5}\right) \quad (A14)$$

Inter-layer couplings:

$$H_{L_{12}} = J_{B_1B_3}\left(\sigma_{B_1}^+\sigma_{B_3}^- + \sigma_{B_1}^-\sigma_{B_3}^+\right) + J_{B_1B_4}\left(\sigma_{B_1}^+\sigma_{B_4}^- + \sigma_{B_1}^-\sigma_{B_4}^+\right) + J_{B_2B_4}\left(\sigma_{B_2}^+\sigma_{B_4}^- + \sigma_{B_2}^-\sigma_{B_4}^+\right) +$$
$$J_{B_2B_5}\left(\sigma_{B_2}^+\sigma_{B_5}^- + \sigma_{B_2}^-\sigma_{B_5}^+\right) \quad (A15)$$

Each bath has its own Hamiltonian and dissipation
$$H_{B_i} = \omega_i\left(\sigma_{B_i}^+\sigma_{B_i}^-\right) \; for \; i = 1, \ldots, 5 \quad (A16)$$

Corresponding Green's Function Equations:

We consider one-excitation basis, ordered as:
$$\mathcal{B} = [|S\rangle, |B_1\rangle, |B_2\rangle, |B_3\rangle, |B_4\rangle, |B_5\rangle]. \quad (A17)$$

Since:
$$\mathbf{H} = \begin{pmatrix} \omega_S & J_{SB_1} & J_{SB_2} & 0 & 0 & 0 \\ J_{SB_1} & \omega_1 - i\gamma_1 & J_{12} & J_{13} & J_{14} & 0 \\ J_{SB_2} & J_{12} & \omega_2 - i\gamma_2 & 0 & J_{24} & J_{25} \\ 0 & J_{13} & 0 & \omega_3 - i\gamma_3 & J_{34} & J_{35} \\ 0 & J_{14} & J_{24} & J_{34} & \omega_4 - i\gamma_4 & J_{45} \\ 0 & 0 & J_{25} & J_{35} & J_{45} & \omega_5 - i\gamma_5 \end{pmatrix} \quad (A18)$$

$$(\omega\mathbf{I} - \mathbf{H}) = \begin{pmatrix} \omega - \omega_S & -J_{SB_1} & -J_{SB_2} & 0 & 0 & 0 \\ -J_{SB_1} & \omega - \omega_1 + i\gamma_1 & -J_{12} & -J_{13} & -J_{14} & 0 \\ -J_{SB_2} & -J_{12} & \omega - \omega_2 + i\gamma_2 & 0 & -J_{24} & -J_{25} \\ 0 & -J_{13} & 0 & \omega - \omega_3 + i\gamma_3 & -J_{34} & -J_{35} \\ 0 & -J_{14} & -J_{24} & -J_{34} & \omega - \omega_4 + i\gamma_4 & -J_{45} \\ 0 & 0 & -J_{25} & -J_{35} & -J_{45} & \omega - \omega_5 + i\gamma_5 \end{pmatrix} \quad (A19)$$

By definition:
$$\mathbf{G}(\omega) = (\omega\mathbf{I} - \mathbf{H})^{-1} \quad (A20)$$

Thus, from the product: $(\omega\mathbf{I} - \mathbf{H})\mathbf{G}(\omega) = \mathbf{I}$, we can obtain several relationships:

For B$_1$:
$$(\omega - \omega_1 + i\gamma_1)G_{B_1S} = J_{SB_1}G_{SS} + J_{B_1B_2}G_{B_2S} + J_{B_1B_3}G_{B_3S} + J_{B_1B_4}G_{B_4S} \quad (A21)$$

For B$_2$:
$$(\omega - \omega_2 + i\gamma_2)G_{B_2S} = J_{SB_2}G_{SS} + J_{B_1B_2}G_{B_1S} + J_{B_2B_4}G_{B_4S} + J_{B_2B_5}G_{B_5S} \quad (A22)$$

For B3:
$$(\omega - \omega_3 + i\gamma_3)G_{B_3S} = J_{B_1B_3}G_{B_1S} + J_{B_3B_4}G_{B_4S} + J_{B_3B_5}G_{B_5S} \quad (A23)$$

For B4:
$$(\omega - \omega_4 + i\gamma_4)G_{B_4S} = J_{B_1B_4}G_{B_1S} + J_{B_2B_4}G_{B_2S} + J_{B_3B_4}G_{B_3S} + J_{B_4B_5}G_{B_5S} \quad (A24)$$

For $B_5$:
$$(\omega - \omega_5 + i\gamma_5)G_{B_5S} = J_{B_2B_5}G_{B_2S} + J_{B_3B_5}G_{B_3S} + J_{B_4B_5}G_{B_4S} \tag{A25}$$

For System:
$$(\omega - \omega_S)G_{SS} = 1 + J_{SB_1}G_{B_1S} + J_{SB_2}G_{B_2S} \tag{A26}$$

In the matrix form:
$$\begin{pmatrix} \omega - \omega_S & -\mathbf{J}_{SB} \\ -\mathbf{J}_{SB} & \mathbf{M} \end{pmatrix} \begin{pmatrix} G_{SS} \\ \mathbf{G_B} \end{pmatrix} = \begin{pmatrix} 1 \\ 0 \end{pmatrix} \tag{A27}$$

$$\mathbf{M}\mathbf{G_B} = \mathbf{J}_{SB}G_{SS} \tag{A28}$$

$$\mathbf{M} = \omega\mathbf{I} - \mathbf{H_B} + i\mathbf{\Gamma} =$$
$$\begin{pmatrix} \omega - \omega_1 + i\gamma_1 & -J_{B_1B_2} & -J_{B_1B_3} & -J_{B_1B_4} & 0 \\ -J_{B_1B_2} & \omega - \omega_2 + i\gamma_2 & 0 & -J_{B_2B_4} & -J_{B_2B_5} \\ -J_{B_1B_3} & 0 & \omega - \omega_3 + i\gamma_3 & -J_{B_3B_4} & -J_{B_3B_5} \\ -J_{B_1B_4} & -J_{B_2B_4} & -J_{B_3B_4} & \omega - \omega_4 + i\gamma_4 & -J_{B_4B_5} \\ 0 & -J_{B_2B_5} & -J_{B_3B_5} & -J_{B_4B_5} & \omega - \omega_5 + i\gamma_5 \end{pmatrix} \tag{A29}$$

$$\mathbf{G_B} = \begin{pmatrix} G_{B_1S} \\ G_{B_2S} \\ G_{B_3S} \\ G_{B_4S} \\ G_{B_5S} \end{pmatrix} \quad (A30) \qquad \mathbf{J}_{SB} = \begin{pmatrix} J_{SB_1} \\ J_{SB_2} \\ 0 \\ 0 \\ 0 \end{pmatrix} \quad (A31)$$

From equation A28:
$$\mathbf{G_B} = \mathbf{M}^{-1}\mathbf{J}_{SB}G_{SS} \tag{A32}$$

Subsituting to equation A27:
$$(\omega - \omega_S)G_{SS} = 1 + \begin{pmatrix} J_{SB_1} & J_{SB_2} & 0 & 0 & 0 \end{pmatrix} \mathbf{G_B} \tag{A33}$$

$$(\omega - \omega_S)G_{SS} = 1 + \mathbf{J}_{SB}^T \mathbf{M}^{-1}\mathbf{J}_{SB}G_{SS} \tag{A34}$$

$$G_{SS} = \frac{1}{(\omega-\omega_S) - \mathbf{J}_{SB}^T \mathbf{M}^{-1}\mathbf{J}_{SB}} \tag{A35}$$

With the self-energy:
$$\boxed{\Sigma_S\omega = \mathbf{J}_{SB}^T \mathbf{M}^{-1}\mathbf{J}_{SB}} \tag{A36}$$

**Phase convention.** In this formulation the complex structure enters through the bath resolvent $\mathbf{M}(\omega) = \omega\mathbf{I} - \mathbf{H_B} + i\mathbf{\Gamma}$, which ensures reciprocity via its Hermitian and anti-Hermitian parts. The system–bath couplings $\mathbf{J}_{SB}$ are taken real, so $\mathbf{J}_{SB}^T = \mathbf{J}_{SB}^\dagger$. If the couplings carry complex phases, one should replace $^T$ by $^\dagger$ throughout; the resulting expressions remain valid and automatically preserve reciprocity.

## Appendix 2 — Self-Energy Derivation for HEOM and TN/TEMPO

This appendix reviews the derivation of effective self-energy expressions used in the Hierarchical Equations of Motion (HEOM) and Tensor-Network (TN/TEMPO) formalisms. These derivations are not original and are well established in the quantum dissipation literature. They are included solely to provide a basis for comparison with the structured-bath formulation introduced in the main text. HEOM and TN/TEMPO reconstruct the bath influence function in distinct mathematical forms—HEOM through exponential expansions of the correlation kernel, TN/TEMPO through chain mappings of the spectral density. Both approximate the same analytic structure that the structured-bath model recovers exactly from a finite, quantized Hamiltonian. This comparison highlights how the structured approach restores architectural reciprocity.

### A. HEOM formulation
Step 1 – Bath correlation expansion.
The bath operator is written as

$$B = \sum_k g_k b_k, \quad H_B = \sum_k \omega_k b_k^\dagger b_k. \qquad (A37)$$

For a thermalized Gaussian bath, the two-time correlation function is expressed as a finite exponential sum

$$C(t) = \langle B(t) B^\dagger(0) \rangle = \sum_j c_j e^{-\gamma_j t}, \qquad (A38)$$

where each coefficient $c_j$ and rate $\gamma_j$ arises from a Padé or Matsubara expansion of the Bose function.

Step 2 – Laplace transform and self-energy.
Applying the unilateral Laplace transform to Eq. (A38) gives

$$\boxed{\Sigma_{\text{HEOM}}(\omega) = \int_0^\infty C(t) e^{i\omega t} \, dt = \sum_j \frac{c_j}{\gamma_j - i\omega}}. \qquad (A39)$$

Each exponential mode represents an auxiliary density operator in the HEOM hierarchy. Eq. (A39) therefore shows that HEOM decomposes the memory kernel into a set of localized exponential channels, each providing a unidirectional decay pathway.

Step 3 – Interpretation.
Because every term in Eq. (A39) corresponds to an independent dissipative pole, the coupling between coherence and loss is reconstructed phenomenologically rather than derived microscopically.
Reciprocity between emission and reabsorption is not imposed at the Hamiltonian level but recovered statistically through convergence of the exponential expansion.

### B. TN/TEMPO formulation
Step 1 – Chain mapping of the spectral density.
A continuous bath with spectral density $J(\omega)$ is discretized by orthogonal polynomial transformation into a semi-infinite chain,

$$H_{\text{chain}} = H_S + \sum_n \epsilon_n b_n^\dagger b_n + \sum_n (t_n b_n^\dagger b_{n+1} + t_n^* b_{n+1}^\dagger b_n) + \lambda(L^\dagger b_0 + b_0^\dagger L). \tag{A40}$$

Step 2 – Recursive elimination of bath sites.
Successively integrating out bath degrees of freedom through Dyson's equation yields a continued-fraction self-energy,

$$\boxed{\Sigma_{\text{TN}}(\omega) = \cfrac{\lambda^2}{\omega - \epsilon_0 - \cfrac{t_0^2}{\omega - \epsilon_1 - \cfrac{t_1^2}{\omega - \epsilon_2 - \cdots}}}}. \tag{A41}$$

Each level in the continued fraction represents a virtual site in the chain. Feedback is encoded through nested energy denominators, not through direct geometric coupling.

Step 3 – Time-nonlocal kernel.
Inverse Fourier transformation of Eq. (A41) produces a convolution kernel,

$$K_{\text{TN}}(t) = \int \frac{d\omega}{2\pi} \Sigma_{\text{TN}}(\omega) e^{-i\omega t}, \tag{A42}$$

which governs the non-Markovian memory in the TN/TEMPO evolution. The kernel decays algebraically with the bandwidth and chain length, reproducing long-tail correlations but without a closed analytic expression for reciprocity.

## C. Comparison with the Structured-Bath Hamiltonian

Step 1 – Unified structure.
Both HEOM and TN/TEMPO approximate the same analytic continuation of the bath correlation function,

$$\Sigma(\omega) = \int_0^\infty C(t) e^{i\omega t}\, dt, \tag{A43}$$

but they differ in how reciprocity enters. HEOM expands $C(t)$ as independent exponential channels, while TN/TEMPO maps it into a tridiagonal chain.

Step 2 – Microscopic derivation in the structured bath.
In the structured-bath Hamiltonian, the self-energy is obtained directly from the finite coupling matrix $J_{ij}$,

$$\boxed{\Sigma_{\text{SB}}(\omega) = V^\dagger (\omega I - H_B)^{-1} V}, \tag{A44}$$

where $V$ collects the system–bath coupling amplitudes. The real and imaginary components of Eq. (A44) arise simultaneously from the same couplings, producing a Hermitian–anti-Hermitian pair that enforces reciprocity at the Hamiltonian level.

Step 3 – Conceptual synthesis.
HEOM and TN/TEMPO reconstruct the environment by fitting its response in abstract spaces. The structured-bath model derives it directly from finite geometry, preserving reciprocity and physical traceability. This distinction defines the practical advantage of the structured approach: it yields a physically transparent self-energy that preserves feedback symmetry and directly connects microscopic topology to macroscopic dissipation.

## Appendix 3 — Derivation of Frequency Correction in the EPR Framework

The Energy Participation Ratio (EPR) method [15] is a well-established framework that begins from a discretized circuit representation, where each electromagnetic element is treated as a finite, localized subsystem contributing to the total stored energy of a given mode. In this respect, the approach parallels the structured-bath construction: both originate from a finite and physically resolved architecture rather than an idealized continuum. However, while the structured-bath model retains a fully quantized description of both the system and its near-field environment, the EPR method applies quantization only after solving the classical field distribution. The environment in EPR remains static and lossless, capturing only the real component of the system's response.

### Step 1: Identification of Participating Elements

Each normal mode $m$ of the circuit possesses a total electromagnetic energy
$$U_{\text{tot},m} = \sum_i U_{i,m}, \quad (A45)$$
where $U_{i,m}$ is the portion stored in element $i$. The energy participation factor
$$p_{i,m} = \frac{U_{i,m}}{U_{\text{tot},m}}, \quad (A46)$$
therefore, quantifies the discrete contribution of each element to mode $m$. This partitioning converts the continuous field description into a finite network of interacting nodes, which is formally similar to the way a structured bath resolves its environment into quantized sites.

### Step 2: Quantization of Circuit Modes

Each classical mode is promoted to a quantum oscillator through
$$\Phi_m \rightarrow \Phi_{m,\text{zpf}}(a_m + a_m^\dagger), Q_m \rightarrow iQ_{m,\text{zpf}}(a_m^\dagger - a_m), \quad (A47)$$
where the subscript zpf denotes the zero-point fluctuations of flux and charge. These represent the finite quantum amplitudes that remain even in the ground state of the oscillator, given by
$$\Phi_{m,\text{zpf}} = \sqrt{\frac{\hbar Z_m}{2}}, Q_{m,\text{zpf}} = \sqrt{\frac{\hbar}{2Z_m}}, \quad (A48\text{-}A49)$$
with $Z_m = \sqrt{L_m/C_m}$ being the characteristic impedance of mode $m$. They satisfy $\Phi_{m,\text{zpf}} Q_{m,\text{zpf}} = \hbar/2$, ensuring that the canonical commutation relation $[\Phi_m, Q_m] = i\hbar$ holds. Physically, these quantities set the natural quantum scale of flux and charge oscillations for each mode. The resulting linear Hamiltonian, representing the harmonic portion of the circuit before nonlinear corrections are included, is
$$H_{\text{lin}} = \sum_m \hbar\omega_m a_m^\dagger a_m, \quad (A50)$$
where $H_{\text{lin}}$ describes the independent harmonic oscillations of all circuit modes obtained from the linearized circuit equations. These modes form the foundation on which the subsequent nonlinear and coupling corrections are applied.

### Step 3: Inclusion of Nonlinear Elements

For a weakly nonlinear Josephson element of energy $E_i$, the potential energy stored in the element is:
$$U_i(\Phi_i) = E_i \left[1 - \cos\left(\frac{2\pi\Phi_i}{\Phi_0}\right)\right] \simeq \frac{E_i}{2}\left(\frac{2\pi\Phi_i}{\Phi_0}\right)^2 - \frac{E_i}{24}\left(\frac{2\pi\Phi_i}{\Phi_0}\right)^4 + \cdots, \quad (A51)$$

where $U_i$ represents the internal (potential) energy, $\Phi_i$ is the local flux through element $i$, and $\Phi_0 = \frac{h}{2e}$ is the flux quantum. The corresponding phase local flux variable is expressed as a superposition of modal fluxes weighted by their participations,

$$\Phi_i = \sum_m p_{i,m} \Phi_{m,\text{zpf}}(a_m + a_m^\dagger). \quad (A52)$$

Substituting this expansion into the energy expression yields the EPR Hamiltonian

$$H_{\text{EPR}} = \sum_m \hbar\omega_m a_m^\dagger a_m + \sum_{i,m,n} E_i\, p_{i,m} p_{i,n}(a_m + a_m^\dagger)(a_n + a_n^\dagger). \quad (A53)$$

Step 4: Extraction of Frequency Correction

Starting from the equation above, the second term couples different modes through their joint participation in the nonlinear elements $i$. To evaluate the frequency shift of each individual mode, we focus on the diagonal terms ($m = n$), which describe self-interaction within mode $m$. Expanding the interacting term: $(a_m + a_m^\dagger)^2 = 2a_m^\dagger a_m + 1 + a_m^2 + a_m^{\dagger 2}.$ (A54)

Under the rotating-wave approximation (RWA), the rapidly oscillating $a_m^2$ and $a_m^{\dagger 2}$ are neglected, giving:

$$H_{\text{diag}} \approx \sum_{i,m} 2E_i p_{i,m}^2 a_m^\dagger a_m + \text{const.} \quad (A55)$$

The constant term merely shifts the energy reference, so the effective Hamiltonian becomes

$$H_{\text{eff}} = \sum_m \hbar\omega_m a_m^\dagger a_m + \sum_{i,m} 2E_i p_{i,m}^2 a_m^\dagger a_m. \quad (A56)$$

Combining these terms leads to a renormalized mode frequency

$$\hbar\omega_m \to \hbar(\omega_m + \Delta\omega_m), \Delta\omega_m = \sum_i E_i p_{i,m}^2. \quad (A57)$$

Hence, the participation of each element $i$ in mode $m$ results in an additive correction to the mode frequency *proportional to* $E_i p_{i,m}^2$. This shift represents the real (Hermitian) part of the self-energy arising from static electromagnetic loading of the nonlinear elements.

Step 5: Relation to an Effective Self-Energy

The Green-function representation of mode $m$ may be written as

$$G_m(\omega) = \frac{1}{\omega - \omega_m - \Sigma_{\text{EPR}}(\omega)} \quad (A58).$$

Within the EPR formulation the correction is *purely real*,

$$\boxed{\Sigma_{\text{EPR}}(\omega) \approx \Delta\omega_m}, \quad (A59)$$

indicating that only the Hermitian (real) component of the self-energy is captured.
The EPR framework therefore maps the geometric distribution of stored energy but omits the imaginary component that would describe feedback or linewidth.

Step 6: Contrast with the Structured-Bath Formalism

The structured-bath Hamiltonian derived in Appendix 1 retains both real and imaginary parts of the self-energy from the same microscopic interactions, producing a complete complex response $\Sigma(\omega) = \text{Re}\Sigma(\omega) + i\text{Im}\Sigma(\omega)$. EPR corresponds to the limit where the bath is frozen into static field geometry. Thus, its participations are fixed and memoryless. The structured-bath approach restores those participations as dynamic, quantized degrees of freedom, allowing coherence, feedback, and gain to emerge naturally from the same microscopic couplings that govern static loading in EPR.

**Appendix 4 — Systematic Parameter Study**

The nine configurations presented here were generated from the same six-node structured-bath Hamiltonian used in Figures 1–3.
The baseline frequency vector
$$\omega = [6.0, 6.5, 6.7, 7.0, 7.2, 7.4] \text{ GHz}$$
represents the system node $S$ and five bath nodes $B_1$–$B_5$.
Each configuration modifies the inter-layer couplings, intra-layer triangular links, and nonlinear gain to trace the amplifier's evolution from near-transparent response to saturated amplification. Low-loss and high-loss limits are represented by $\gamma = [10^{-3}, 2 \times 10^{-2}]$.

Unless otherwise specified, all parameter sweeps are performed in the unpumped regime (P=0), corresponding to the passive structured amplifier baseline. The pump is activated only during the dedicated pump-sweep analysis, where P is continuously increased to reveal the transition from passive response to active gain.

**Passive Structured Response (C1–C3)**
C1 serves as a weak passive baseline rather than a perfectly isolated case.
Minimal but finite couplings ($J_{SB} = 0.05, J_{B1B2} = 0.03, J_{L1L2} = J_{L2}^{\text{int}} = 0.03$) allow a faint coherent trace, ensuring continuity with the structured regimes that follow. modest internal bridge within the first-layer network $L_1$, and begins to activate light $L_1 \rightarrow L_2$ links, producing a shallow structured response. C3 strengthens both the $L_1 \rightarrow L_2$ and intra-layer couplings, converting the initially flat spectrum into one displaying diagonal ridges and alternating bright–dark bands—signatures of passive hybridization and energy exchange between layers.

This diagonal structure only emerges in C3, where $J_{L1L2}$ crosses the threshold needed to support interlayer coherence routing. In C1 and C2, the spectrum remains flat or weakly structured, as $J_{L1L2}$ is too small to sustain frequency-dependent hybridization. By contrast, C3 activates multiple $L_1 \rightarrow L_2$ links simultaneously, allowing coherence to bend across frequency and imprint alternating bright–dark bands. This bending is not pump-driven but arises purely from passive architectural routing, confirming $J_{L1L2}$ as the key control knob for spectral curvature in the absence of gain.

Interestingly, even in C1, the spectrum shows a faint diagonal trace across frequency, reflecting minimal but nonzero interlayer coherence. This confirms that passive routing is architecturally seeded from the outset, with $J_{L1L2}$ and $J_{SB}$ jointly enabling shallow coherence bending. Though not strong enough to form full ridges, this early structure ensures continuity with the hybridized regimes that follow.

In the weak coupling regime of C1–C2, where $J_{L1L2}$ is too small to sustain robust interlayer coherence, $J_{SB}$ effectively substitutes as the dominant routing channel. The bath becomes the primary readout surface, registering faint diagonal traces that bypass the fragmented internal network. This substitution confirms $J_{SB}$ as the architectural fallback for spectral imprinting when internal coherence is underdeveloped.

However, despite this routing role, the overall gain at $B_3$ remains low. Without nonlinear amplification or strong $J_{L1L2}$ coherence, $J_{SB}$ alone cannot sustain high-intensity readout. The bath registers only shallow traces, confirming that passive routing enables visibility but not amplification.

**Driven Amplification and Hybrid Breathing (C4–C6)**
Introducing a finite nonlinear-gain parameter transforms the passive network into an active amplifier characterized by a bright, triangular gain tongue in the spectral maps (see arrows). C4 applies a +15 MHz detune on $B_3$ with moderate drive ($g_{\text{NL}} = 0.20$), initiating the onset of coherent amplification. C5 enhances coupling modulation at fixed frequencies, strengthening the hybrid breathing pattern while keeping the spectral center near 6.6–6.8 GHz. C6 combines asymmetric system–bath coupling ($J_{SB1} \neq J_{SB2}$) with mixed detunes across $B_1, B_2, B_3$. The resulting spectra show directional coherence flow between $L_1$ and $L_2$ and a pronounced breathing rhythm between inner and outer layers, demonstrating that coherence can be steered by coupling asymmetry as effectively as by pump power. Notably, while C4 and C5 exhibit upward ridge migration with increasing pump power, C6 breaks this trend: the ridge folds back into the lower band at higher pump levels, revealing bidirectional breathing and nonlinear redistribution within the gain tongue.

**Memory and Saturation (C7–C9)**
C7 staggers the triangular layer by detuning $B_3$ upward (+10 MHz) and $B_5$ downward (−10 MHz), revealing internal energy recycling and fine interference fringes within the outer bath.
C8 shifts the quartet $S, B_1, B_2, B_3$ upward by +20 MHz to achieve phase-matched amplification, yielding the cleanest, narrowest ridge across all configurations. Finally, C9 represents the saturation regime, combining stronger couplings with a nonlinear gain of 0.55. The gain tongue broadens significantly with increasing pump power, forming a high-intensity coherence plateau across multiple frequency bands. Unlike the triangular tongues in C4–C6, which exhibit directional breathing and frequency switching, C9 suppresses this structure and completes the transition from modular amplification to full spectral saturation.

Table A4 Simulation parameters for the nine configurations

| ID | Description | Freqs [GHz] (S,$B_1$,$B_2$,$B_3$,$B_4$,$B_5$) | $J_{SB}$ | $J_{B_1B_2}$ | $J_{L_1L_2}$ | $J_{L_2}$ | nonlinear gain |
|---|---|---|---|---|---|---|---|
| C1 | Transparent baseline | [6.0, 6.5, 6.7, 7.0, 7.2, 7.4] | [0.05, 0.05] | 0.03 | [0.03, 0.03, 0.03, 0.03] | [0.03, 0.03, 0.03] | 0.00 |
| C2 | Weak structure ($L_1$ only) | same | [0.10, 0.10] | 0.20 | [0.10, 0.08, 0.00, 0.00] | [0.03, 0.03, 0.03] | 0.00 |
| C3 | Structured passive ($L_1 \to L_2$) | same | [0.16, 0.16] | 0.30 | [0.35, 0.30, 0.20, 0.20] | [0.08, 0.06, 0.05] | 0.00 |
| C4 | Pump onset (+15 MHz $B_3$) | [6.0, 6.5, 6.7, 7.015, 7.2, 7.4] | [0.26, 0.24] | 0.40 | [0.45, 0.40, 0.35, 0.35] | [0.10, 0.08, 0.07] | 0.20 |
| C5 | Pump (coupling modulation) | [6.0, 6.5, 6.7, 7.0, 7.2, 7.4] | [0.26, 0.26] | 0.45 | [0.50, 0.45, 0.40, 0.40] | [0.12, 0.10, 0.08] | 0.20 |
| C6 | Asymmetric SB + mixed detune | [6.0, 6.49, 6.71, 7.015, 7.20, 7.40] | [0.32, 0.16] | 0.50 | [0.50, 0.42, 0.36, 0.36] | [0.12, 0.09, 0.08] | 0.30 |
| C7 | Triangle imbalance | [6.0, 6.5, 6.7, 7.010, 7.200, 6.990] | [0.22, 0.22] | 0.40 | [0.42, 0.38, 0.28, 0.28] | [0.14, 0.02, 0.10] | 0.00 |
| C8 | Phase-matched (+20 MHz S,$B_1$,$B_2$,$B_3$) | [6.020, 6.520, 6.720, 7.020, 7.200, 7.400] | [0.24, 0.24] | 0.45 | [0.48, 0.44, 0.34, 0.34] | [0.10, 0.08, 0.06] | 0.00 |
| C9 | Max gain and saturation | [6.0, 6.505, 6.705, 7.035, 7.230, 7.410] | [0.32, 0.32] | 0.55 | [0.60, 0.55, 0.50, 0.50] | [0.16, 0.14, 0.12] | 0.55 |

## CASE1

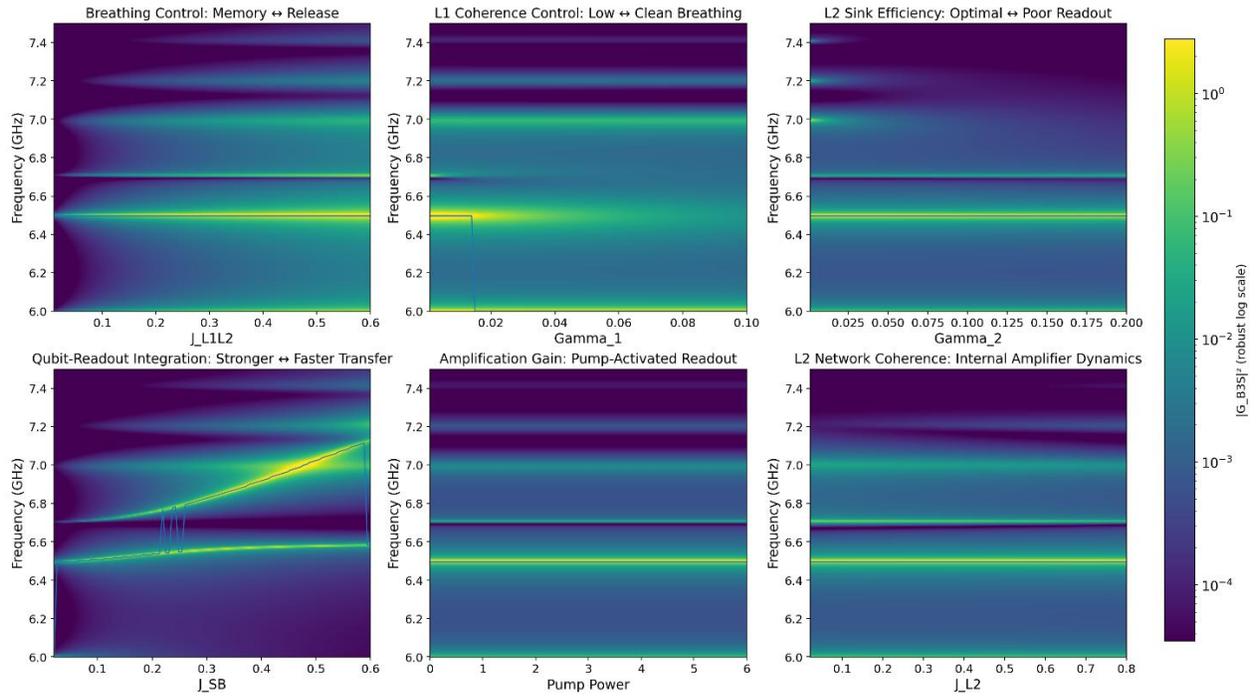

## CASE2

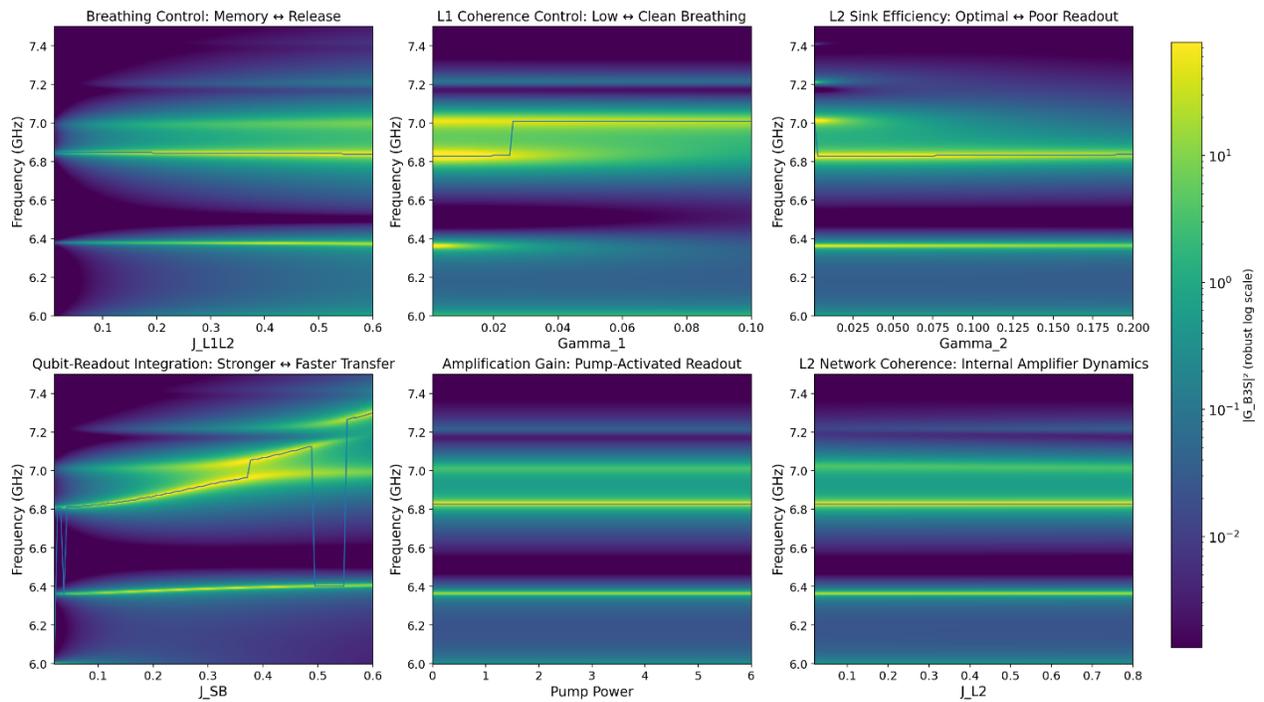

CASE3

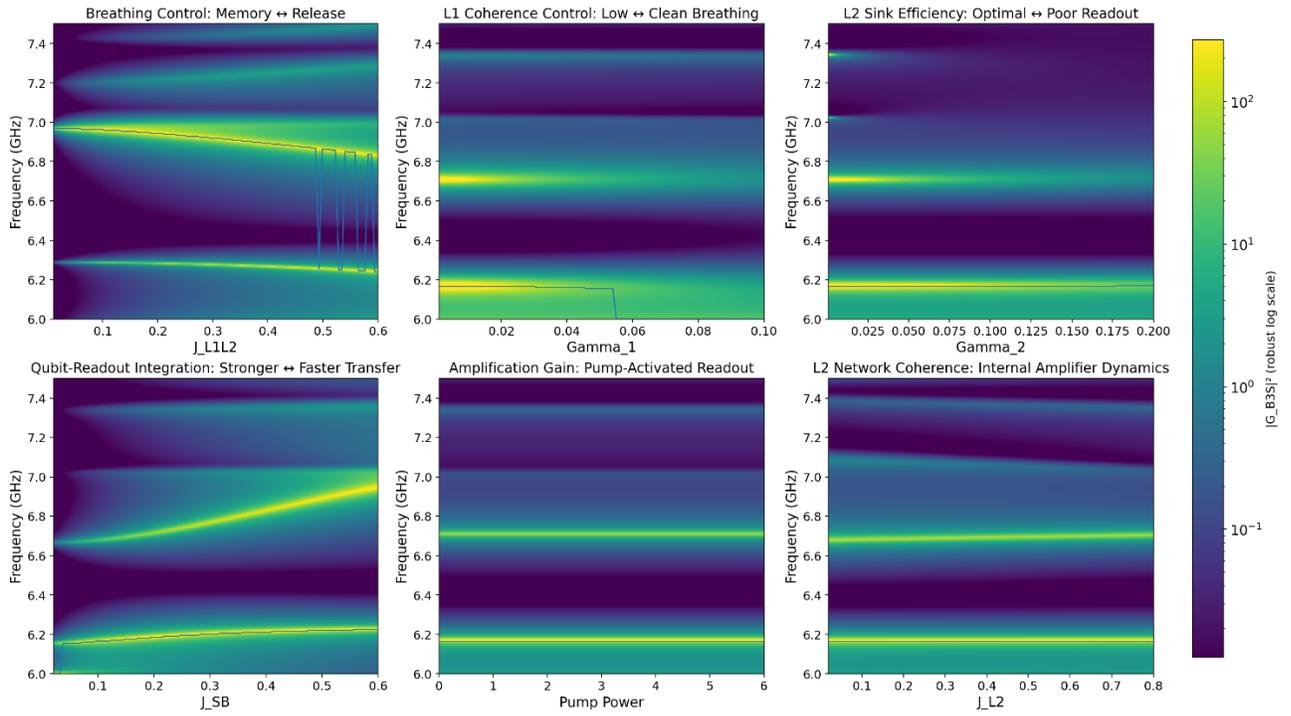

CASE4

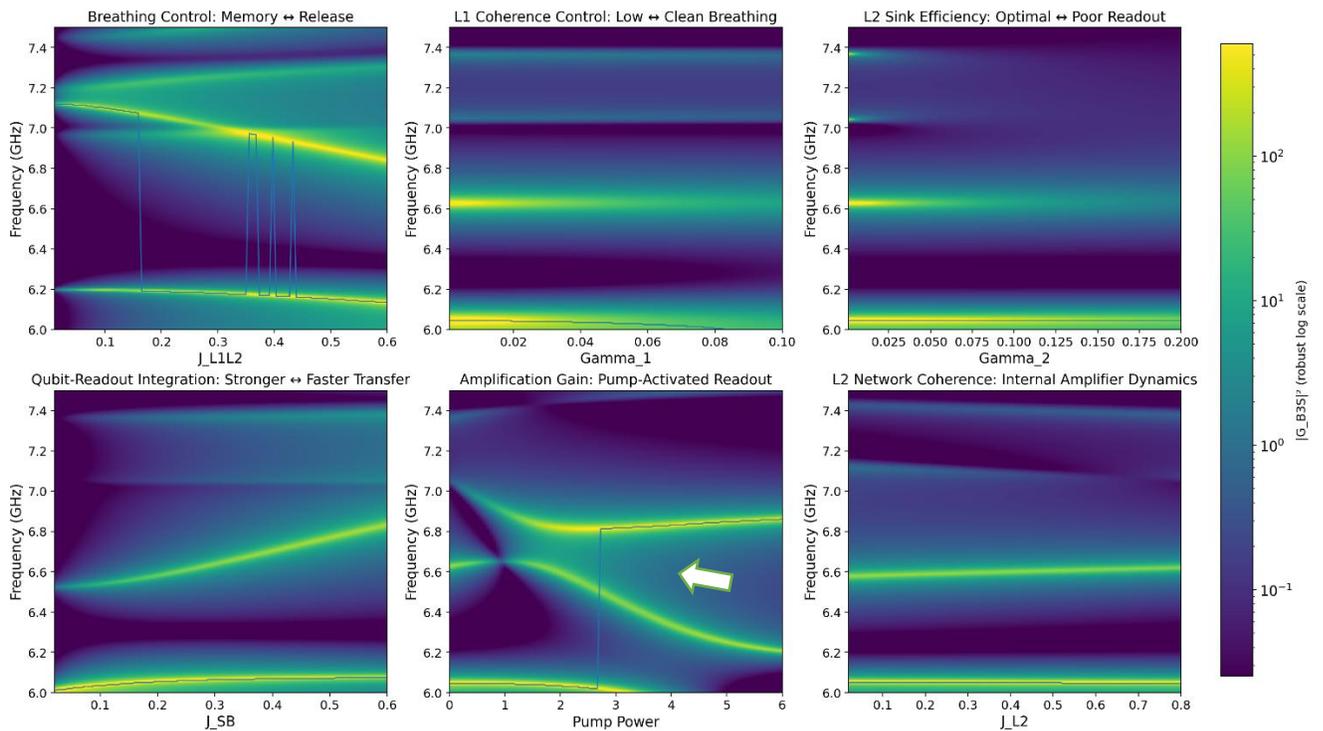

## CASE5

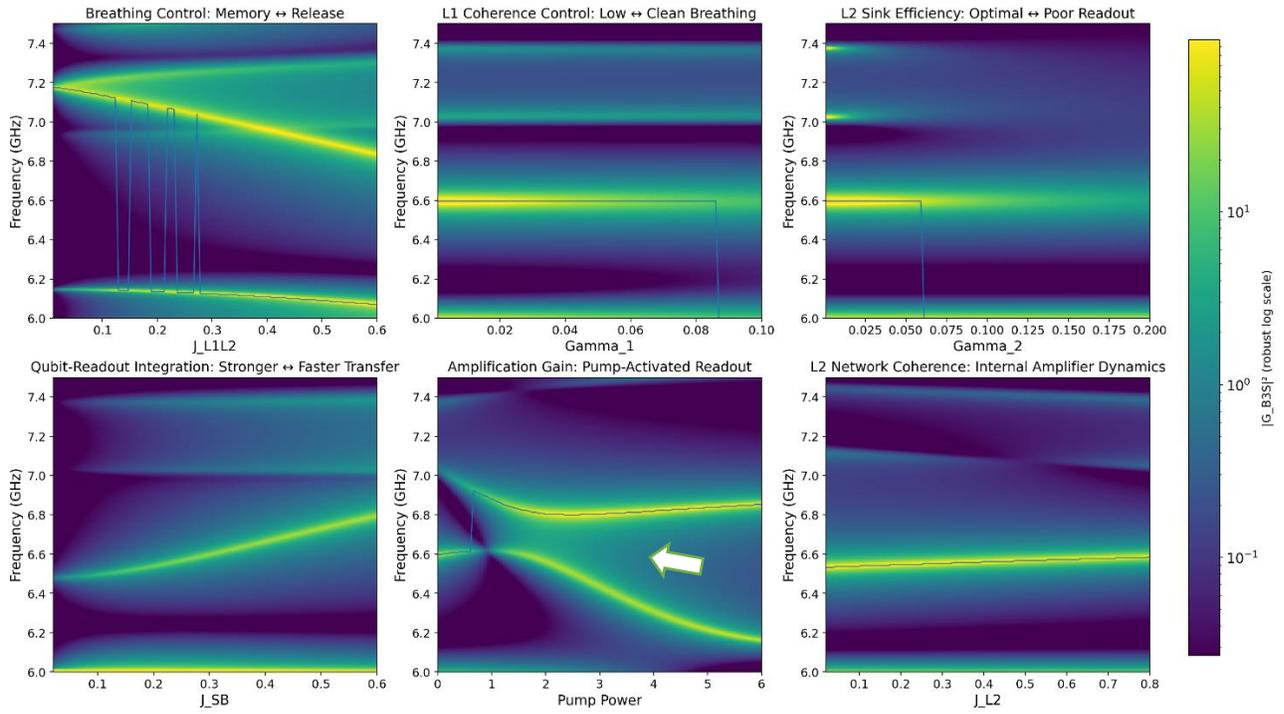

## CASE6

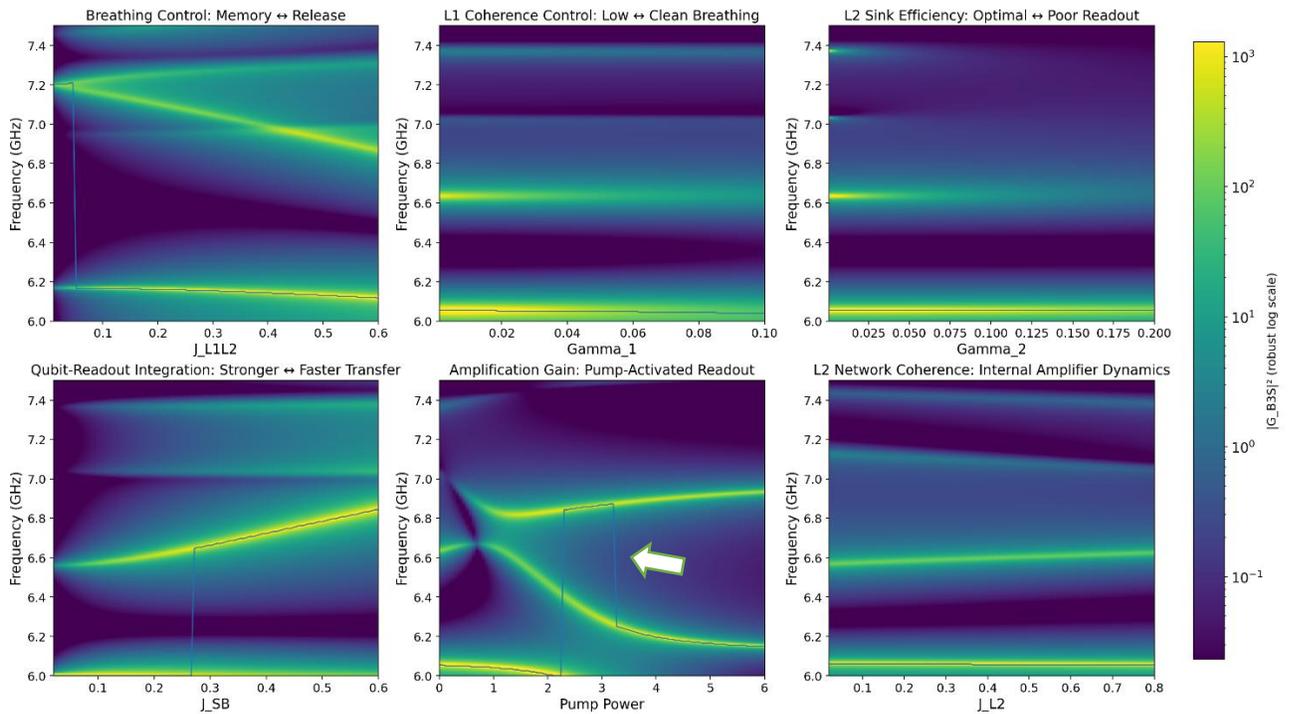

## CASE7

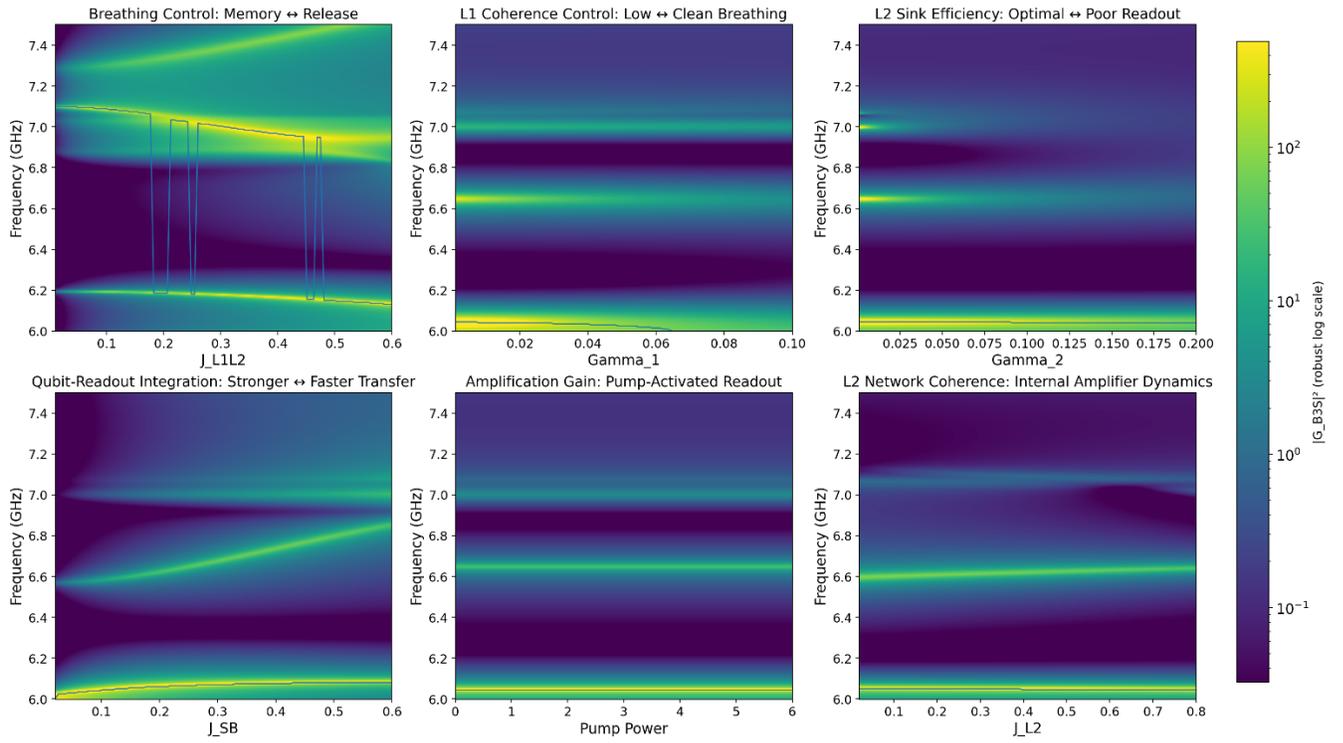

## CASE8

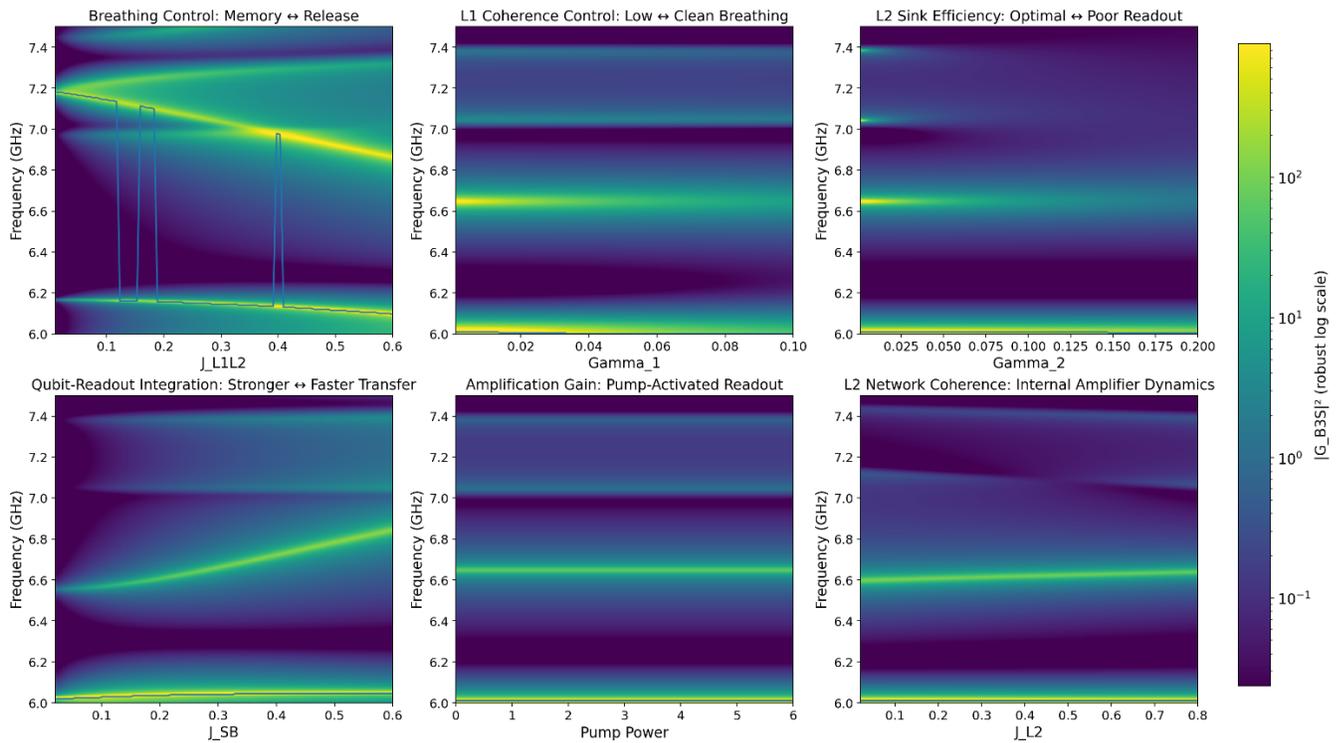

CASE9

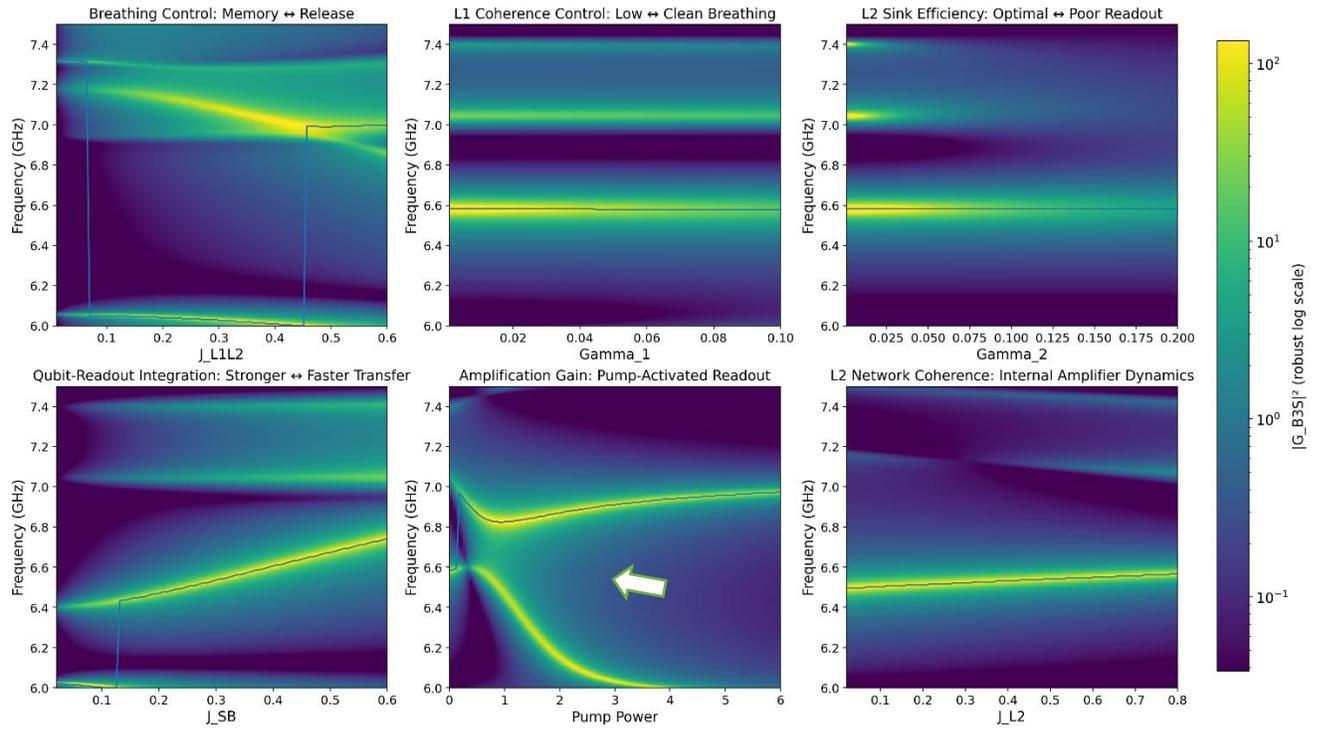